\documentclass[11pt, a4paper]{article}
\pdfoutput=1

\usepackage{jheppub} 

\usepackage{amsmath}
\usepackage{amsfonts}
\usepackage{amssymb}
\usepackage{graphicx}
\graphicspath{{./figs/}}
\usepackage{mathrsfs}
\usepackage{latexsym}
\usepackage{color}
\usepackage{float}
\usepackage[bottom]{footmisc}
\usepackage{slashed, cancel}
\usepackage{hyperref}
\usepackage{bbold}

\usepackage[font=footnotesize,labelfont=bf]{caption} 

\numberwithin{equation}{section}

\definecolor{rossos}{rgb}{0.8,0.2,0.3}
\definecolor{bluscuro}{rgb}{0.15, 0.2, .85}
\definecolor{bluchiaro}{cmyk}{1,.3,0.,0.1}

\definecolor{orange}{rgb}{1,0.5,0}
\definecolor{blue}{rgb}{0,0,1}

\hypersetup{colorlinks, citecolor=bluscuro, linkcolor=bluscuro, urlcolor=bluscuro}

\makeatother   

\def\mx{m_{\rm DM}}
\def\gx{g_{\rm DM}}

 \def\be   {\begin{equation}}   \def\ee   {\end{equation}}
 \def\ba   {\begin{array}}      \def\ea   {\end{array}}
 \def\bea  {\begin{eqnarray}}   \def\eea  {\end{eqnarray}}
 \def\bean {\begin{eqnarray*}}  \def\eean {\end{eqnarray*}}

\title{Mapping monojet constraints onto Simplified Dark Matter Models}

\author[a]{Thomas Jacques,}
\author[b]{Karl Nordstr\"{o}m}

\affiliation[a\,]{D\'epartement de Physique Th\'eorique \& Center for Astroparticle Physics,\\
~\, Universit\'e de Gen\`eve, Quai E.\ Ansermet 24, 1211 Gen\`eve 4, Switzerland}
\affiliation[b\,]{SUPA, School of Physics and Astronomy, University of Glasgow,
Glasgow G12 8QQ, Scotland, UK}

\emailAdd{thomas.jacques@unige.ch}
\emailAdd{k.nordstrom.1@research.gla.ac.uk}

\date{\today}                                           

\abstract{
The move towards simplified models for Run II of the LHC will allow for stronger and more robust constraints on the dark sector. 
However there already exists a wealth of Run I data which should not be ignored in the run-up to Run II. 
Here we reinterpret public constraints on generic beyond-standard-model cross sections to place new constraints on a simplified model. We make use of an ATLAS search in the monojet $+$ missing energy channel to constrain a representative simplified model with the dark matter coupling to an axial-vector $Z'$. 
We scan the entire parameter space of our chosen model to set the strongest current collider constraints on our model using the full 20.3 fb$^{-1}$ ATLAS 8 TeV dataset and provide predictions for constraints that can be set with 20 fb$^{-1}$ of 14 TeV data.
Our technique can also be used for the interpretation of Run II data and provides a broad benchmark for comparing future constraints on simplified models.
}
\begin{document}

\hfill

\maketitle

\section{Introduction}

In recent years Effective Field Theories (EFTs) have become a popular framework with which to constrain the dark sector at the LHC \cite{Cao:2009uw,Beltran:2010ww,Bai:2010hh,Goodman:2010ku, Cheung:2010ua,Zheng:2010js,Cheung:2011nt,Rajaraman:2011wf,Lowette:2014yta,Yu:2011by,Goodman:2010yf,Fox:2011pm}. In the simplest cases, the dark couplings and mediator masses are combined into a single effective energy scale, $\Lambda$,\footnote{Also called $M_\star$ in the literature.} leaving this and the dark matter mass, $\mx$, as the only free parameters for each effective operator. EFT constraints have the advantage of being relatively model-independent, allowing constraints to be placed across a broad range of models and parameters. In addition they facilitate an easy comparison with direct detection experiments via the shared energy scale $\Lambda$. However it is now clear that EFTs must be used with extreme care at LHC energies, where the energy scale is large enough that the approximations used in the construction of EFTs can not be assumed to be valid. At these energies and luminosities, the energy carried by the mediator is usually larger than the mediator mass, violating the EFT approximations, except in the case of large mediator masses or for dark-sector couplings approaching the perturbativity limit \cite{Busoni:2013lha,Busoni:2014sya,Busoni:2014haa, Buchmueller:2013dya,Goodman:2010yf,Fox:2011pm,Fox:2011fx,Goodman:2011jq,Shoemaker:2011vi,Fox:2012ee,Weiner:2012cb}. Depending on the mass and width of the mediator, this can lead to EFT constraints that are either stronger or weaker than the constraints would be on a UV-complete model, reducing their utility and making their validity questionable. 

One solution is to rescale EFT constraints, by truncating the simulated signal such that only events for which the EFT approximation are valid are used to derive constraints \cite{Busoni:2014sya,ATL-PHYS-PUB-2014-007,Aad:2015zva}. This weakens constraints but at the same time makes them substantially more \emph{robust}, which is critical when considering bounds on beyond-standard-model parameters.
Whilst this technique has the advantage of maintaining some of the elegance of EFTs, it also has the serious disadvantage that it does not make full use of all potential signal events available in a UV complete model and so does not address the region of parameter space where EFT constraints are too weak. To constrain this region we need to consider models where the mediator can be resolved. On the other hand, the parameter space of full, well-motivated models such as supersymmetry \cite{Chung:2003fi} or extra dimensions \cite{ArkaniHamed:1998rs} is broad, and by focusing solely on such models we run the risk of missing more generic signatures of the dark sector. 

Hence, the usage of simplified models is now advocated by a number of groups \cite{Abdallah:2014dma,Malik:2014ggr,Buchmueller:2014yoa,Alves:2011wf,Harris:2014hga,Buckley:2014fba}. Here we will use publicly available ATLAS constraints on the monojet $+$ missing energy channel to constrain a simplified model with dark matter coupling to the standard model via exchange of an axial-vector $Z'$ mediator. The original search was used to constrain EFTs, however the same data and analysis can be used to constrain a simplified model of choice through the model-independent limit on the visible cross section contribution from beyond-standard-model processes. Such a reanalysis only requires simulation of the signal in the new model for each point in parameter space. 

 Simplified models have the advantage of a relatively small set of free parameters, and do not encounter the same validity problems as EFTs. However, the parameter space is still larger than for EFTs, which often necessitates arbitrary choices for one or more parameters in order to constrain the remaining free parameters. Here we will instead leave the dark matter mass, mediator mass, and coupling strength all as free parameters which we scan over and constrain in contours. We derive limits using the full ATLAS 8 TeV dataset \cite{Aad:2015zva} and predict the limits that can be set with 20 fb$^{-1}$ of 14 TeV data.

In Section~\ref{model}, we outline the choice of simplified model that we will be constraining. In Section~\ref{method}, we describe our technique for converting the model-independent constraints on the visible monojet cross section 
into constraints on this simplified model. In Sections~\ref{8tev} and \ref{14tev} we present our results, before we give our concluding remarks in Section~\ref{conclusion}. The Appendix contains validations of our limit-setting and various approximations used, and a discussion of the relic density calculation.

\section{Model\label{model}}

We consider a widely-used benchmark simplified model where Dirac DM interacts with the SM via a $Z'$-type mediator. This is described by the following Lagrangian interaction term:
\begin{equation}
\mathcal{L} = -\sum_f Z'_\mu  [\bar q \gamma^\mu (g^V_q - g^A_q \gamma_5) q] - Z'_\mu \
[\bar \chi \gamma^\mu(\gx^V - \gx^A \gamma_5)\chi],
\label{lagr}
\end{equation}
where $g_i^V,\,g_i^A$ are respectively the vector and axial-vector coupling strengths between the mediator and quarks ($i=q$) and DM ($i=$DM).  This is a well-motivated simplified model that has been studied extensively, including searches by CMS \cite{Chatrchyan:2013qha} and ATLAS \cite{Aad:2012hf}, and numerous other groups both in the UV complete and EFT limits, e.g. Ref.~\cite{Gershtein:2008bf,Buchmueller:2014yoa,Harris:2014hga}. It is part of the wider family of dark $Z'$ portal models which have been studied previously in e.g. \cite{An:2012va,Alves:2013tqa,Alves:2015pea,Lebedev:2014bba,Frandsen:2012rk}.
%
The LHC is relatively insensitive to the mixture of Vector/Axial-vector couplings \cite{Aad:2015zva}, however this ratio has a large effect on the sensitivity of direct detection experiments to this model. A vector coupling induces a spin-independent (SI) WIMP-nucleon scattering rate, while an axial-vector coupling induces a spin-dependent (SD) rate \cite{DelNobile:2013sia}. Current bounds on SI interactions are much stronger than those on SD, to the point where direct detection constraints are generally stronger than LHC constraints on models with pure vector couplings, and vice-versa for pure axial-vector couplings, as seen in e.g. Ref.~\cite{Aad:2014tda}. For this reason we consider a pure axial-vector coupling, setting  $\gx^V = g_q^V = 0$, and defining $\gx \equiv \gx^A$, $g_q \equiv g_q^A$. 
For simplicity, as is common, we assume that the quark-mediator coupling $g_q$ is the same for each species of quark. We require that $\gx$, $g_q \leq 4\pi$ individually in order for the couplings to remain in the perturbative regime, however in practice limiting $\Gamma_\mathrm{OS}/M$ ends up constraining our parameter space further.


For the model we consider, the total on-shell width of the axial-vector mediator is given by:
\begin{equation}
\Gamma_\mathrm{OS} =  \frac{\gx^2 M (1 - 4\mx^2/M^2)^{3/2}}{12 \pi} \Theta(M - 2 \mx) +  \sum_q\frac{g_q^2 M (1 - 4m_q^2/M^2)^{3/2}}{4 \pi} \Theta(M - 2 m_q),
\end{equation}
where $M$ is the mediator mass. With the assumption that $g_q$ is equal for each flavor of quark this width can become very large, for example rising above $\Gamma_\mathrm{OS} \sim M$ at $g_q = \gx \approx 1.45$ when $g_q = \gx$. This width assumes no additional decay channels aside from quarks and DM, however it is conceivable that the mediator could decay to standard model leptons or additional dark sector particles. Given that the the structure of a possible dark sector is unknown and that couplings to leptons are more appropriately constrained by searches for dilepton resonances \cite{Aad:2014cka,Khachatryan:2014fba}, we confine ourselves to the more `minimal' model where the mediator couples only to quarks and DM. For a study of how the limits change when the width is manually made larger (without considering specific additional decay modes) see \cite{Harris:2014hga}.%

\begin{figure}[h]
\centering
 \includegraphics[width=.8\textwidth]{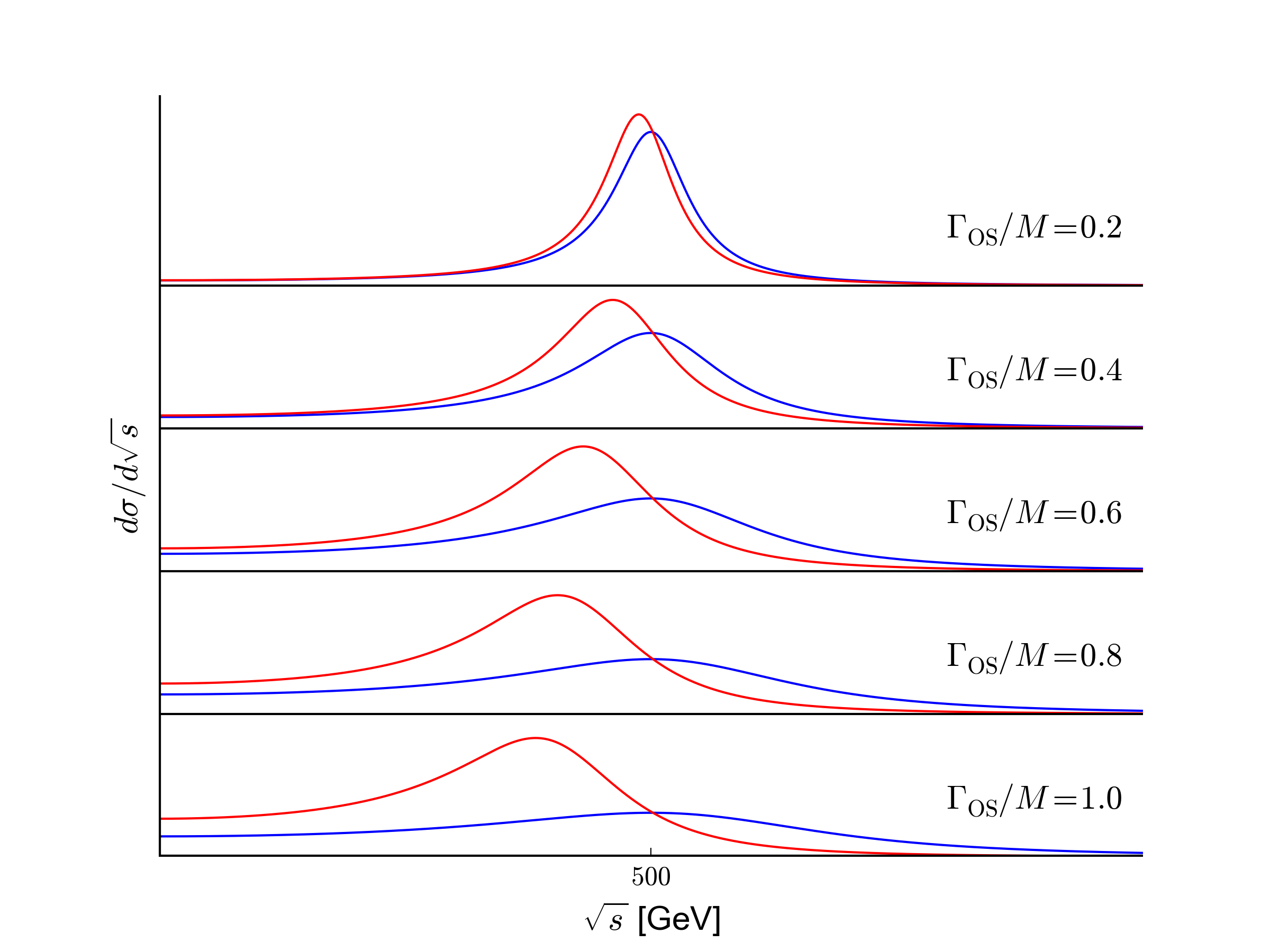}
 \caption{Line shapes for our Breit-Wigner propagator \ref{BW} (blue) and a kinetic propagator with the replacement $\Gamma_\mathrm{OS} M \to s \Gamma(s)/M$ in the imaginary part of the self-energy (red) for various values of $\Gamma_\mathrm{OS}/M$. The normalisation is arbitrary and differs between the plots to allow for a straightforward shape comparison, and both axes are linear. $M=500$ GeV and $\mx=100$ GeV which corresponds to a typical parameter space point.}
 \label{comparison}
 \end{figure}

In the event generation we will use a Breit-Wigner form for the $Z'$ propagator with constant on-shell widths:
\begin{equation}
 \Delta_{Z'}(s, M, \Gamma) = \frac{1}{s - M^2 + i M \Gamma_\mathrm{OS}}
 \label{BW}
\end{equation}
This is questionable from a first-principle Quantum Field Theory perspective \cite{Goria:2011wa,Englert:2015zra}: it amounts to a replacement of $s \Gamma(s)/M \to \Gamma(s) M$ in the imaginary part of the Dyson resummed self-energy, which clearly can only be motivated for small widths for which $s \sim M^2$. Additionally we identify $\Gamma(s)$ as its on-shell value $\Gamma_\mathrm{OS}$; This is formally correct in our tree-level context, but can only be expected to be a good approximation when $\Gamma_\mathrm{OS}/M \ll 1$. A completely consistent theoretical treatment of our model in the part of parameter space where we can't assume that $\Gamma_\mathrm{OS}/M \ll 1$ hence requires a more careful consideration of the propagator than we employ in our simulation, and we therefore use cross section reweighting for the parts of parameter space where we are not conservative to correct for this. In Figure~\ref{comparison} we show a comparison of line shapes for various values of $\Gamma_\mathrm{OS}/M$ with fixed $M, \mx$ using the naive Breit-Wigner shape we employ and a fully kinetic propagator, which reflects that our simulation will become increasingly poor as $\Gamma_\mathrm{OS}/M$ increases, as shown previously in \cite{An:2012va}. There are also further corrections which can not be included by using a kinetic width at tree level as we do here, but these are subleading and this treatment should be sufficient for our purposes.

We have estimated the ratio of final cross sections using a Breit-Wigner propagator and a kinetic one by convolving them with parton density functions (PDFs) and requiring that we can produce a $\chi \bar{\chi}$ pair on-shell and found that within this approximation the choice of propagator makes for modest differences for most of our parameter space, with large effects in specific regions: for $\mx \ll M$ and $\Gamma_\mathrm{OS}/M \gtrsim 0.5$ our propagator gives a cross section that can be as much as 50\% too low compared to a kinetic propagator. In the other extreme end where $\mx \gtrsim M$ and $\Gamma_\mathrm{OS}/M \gtrsim 0.5$ we see an opposite effect where our cross section can be several times too large. Since our propagator in this region is not conservative we will reweight the visible cross section we get at the end of our simulation whenever $\sigma_\mathrm{Breit-Wigner} > \sigma_\mathrm{kinetic}$ using the ratio of the two (where these cross sections are estimated as described above) to allow our limits to remain robust.

\section{Reinterpreting Monojet Constraints\label{method}}

 Our signal prediction is obtained by implementing the model in the \textsc{FeynRules} \cite{Alloul:2013bka} and \textsc{MadGraph5\_aMC@NLO} 2.1.2 \cite{Alwall:2014hca} framework to generate leading order (LO) parton level events using the NNPDF2.3 LO PDFs \cite{Ball:2012cx}. These are matched to \textsc{Pythia} 8.185 \cite{Sjostrand:2007gs} using the MLM algorithm with a matching scale of $80$ GeV\footnote{Chosen to correspond to the matching scale used in the original ATLAS EFT interpretation.} for showering and hadronisation using tune 4C. We generate $\chi \bar\chi$ +  0, 1, and 2 jets in the matrix element before matching to the parton shower. We use the default \textsc{MadEvent} factorization and renormalization scales ($\mu_{R,F}$) which in this case both are approximately the transverse mass of the $\chi \bar{\chi}$ system. Our approach only makes leading order + parton shower (LOPS) predictions compared to the next-to-leading order + parton shower (NLOPS) predictions used in a similar study of CMS results \cite{CMS-PAS-EXO-12-048} in \cite{Buchmueller:2014yoa}, which means we suffer from larger theoretical uncertainty due to scale dependencies which we can attempt to estimate by varying our choice of $\mu_{R,F}$ by a factor of two. This shows a weak dependence on the choice of scales of $_{-5\%}^{+10\%}$ for a few representative choices of $M, \mx$ which is clearly not a realistic estimate of the uncertainty: previous studies \cite{Haisch:2012kf,Fox:2012ru,Haisch:2013ata} with other choices of scales have found fixed-order NLO corrections ranging from $\sim 20-40\%$.
We do however note that based on the results in \cite{Haisch:2013ata}, we expect fixed-order NLO corrections to ultimately be modest after matching to a parton shower and applying the ATLAS monojet analysis cuts since the parton shower dilutes differences, helped by the loose cuts on additional jets. As such they should have a limited impact on our quantitative results and be negligible for qualitative results. 

We analyze the generated events using the \textsc{ATOM} framework \cite{ATOMpaper,Papucci:2014rja} based on Rivet \cite{Buckley:2010ar}. We first divide the final state into topological clusters and find jets with the anti-$k_t$ algorithm \cite{Cacciari:2008gp} using $R = 0.4$ in \textsc{FastJet} \cite{Cacciari:2011ma}. We then perform a smearing of the $p_\textrm{T}$ of these jets based on typical values for the ATLAS detector, leaving the $E_\textrm{T}^\textrm{miss}$ unsmeared\footnote{We are not aware of any ATLAS $E_\textrm{T}^\textrm{miss}$ smearing values which could be unambiguously applied to our case, based on the results in \cite{ATLAS-CONF-2013-082} we expect the plateau to have been reached for all our signal regions however.}. Our procedure has been validated by recreating the limits set on an EFT operator by ATLAS and the results of this validation is given in Appendix~\ref{validation}.

Some past constraints on simplified models have used a fixed benchmark width. In this case, the cross section is only sensitive to the \emph{product} $\gx \cdot g_q$ and not to the couplings individually; Further, this easily factorises out, 
\begin{equation}
d\sigma(\gx, g_q) = (\gx \cdot g_q)^2 d\sigma(\gx = g_q = 1),
\end{equation}
which simplifies the analysis since the coupling affects only the magnitude of the signal, not the spectral shape.
 Including the physical width complicates things, since now both the magnitude and signal spectrum have a dependence on both $\gx \cdot g_q$ and $g_q/\gx$. 
This results in necessary complications if one wants to present 2D contour limits on $\gx \cdot g_q$ when the width is known. 
However it is possible to make an approximation for the cross section in the resonant region as $\sigma \propto g_q^2 \gx^2/\Gamma_\mathrm{OS}$ (for fixed $M, \mx$) and $\sigma \propto g_q^2 \gx^2$ for the off-shell region, which allows us to set limits on $\gx \cdot g_q$ while avoiding a scan in this dimension, leaving only $M - \mx$ as free parameters for any given choice of $g_q/\gx$. This approximation should work well for the part of parameter space where $\Gamma_\mathrm{OS} \ll M$ \cite{Han:2005mu,Fox:2011fx}, and we present a full study of this approximation (including the effect of the mediator shape reweighting) in Appendix~\ref{reweighting} which further motivates restricting the parameter space to $\Gamma_\mathrm{OS}/M < 0.5$.

We also include relic density constraints by finding out which parts of our parameter space would result in a larger relic abundance than observed experimentally. The details of this calculation are described in Appendix~\ref{relics}.

\section{20.3 fb$^{-1}$ 8 TeV Limits \label{8tev}}

 We make use of 90\% C.L. limits set using the full ATLAS 8 TeV dataset \cite{Aad:2015zva} which should allow us to set the strongest collider constraints yet on our model due to improvements in the selection and the use of more signal regions compared to the full dataset analysis by CMS \cite{Khachatryan:2014rra}. The analysis defines signal regions based on $E_{\mathrm{T}}^{\mathrm{miss}}$ and initially only requires the leading jet $p_T^{j1} > 120$ GeV with the additional requirement that $2 \cdot p_T^{j1} > E_{\mathrm{T}}^{\mathrm{miss}}$ for the harder signal regions, and does not veto events due to additional jets as long as $\Delta \phi(\mathrm{jet}, E_{\mathrm{T}}^{\mathrm{miss}}) > 1.0$. The signal regions are given in Table~\ref{full8tevtable}.

\begin{table}[h]
\centering
\begin{tabular}{|c|c|c|c|c|c|c|c|c|c|}
\hline
Signal Region & SR1 & SR2 & SR3 & SR4 & SR5 & SR6 & SR7 & SR8 & SR9 \\
\hline
$E_{\mathrm{T}}^{\mathrm{miss}}$ $\&$ $2 \cdot p_T^{j1}$ $>$ [GeV]& 150 & 200  & 250 & 300 & 350 & 400 & 500 & 600 & 700 \\
ATLAS $\sigma_\textrm{vis}^\textrm{90\% CL}$ [fb] & 599 & 158 & 74 & 38 & 17 & 10 & 6.0 & 3.2 & 2.9 \\
\hline
\end{tabular}
\caption{Signal region definitions in the full dataset 8 TeV analysis and 90 \% CL exclusion limits on the visible cross section from BSM contributions.}
\label{full8tevtable}
\end{table}

 \begin{figure}[ht]
\centering
\includegraphics[width=0.49\textwidth]{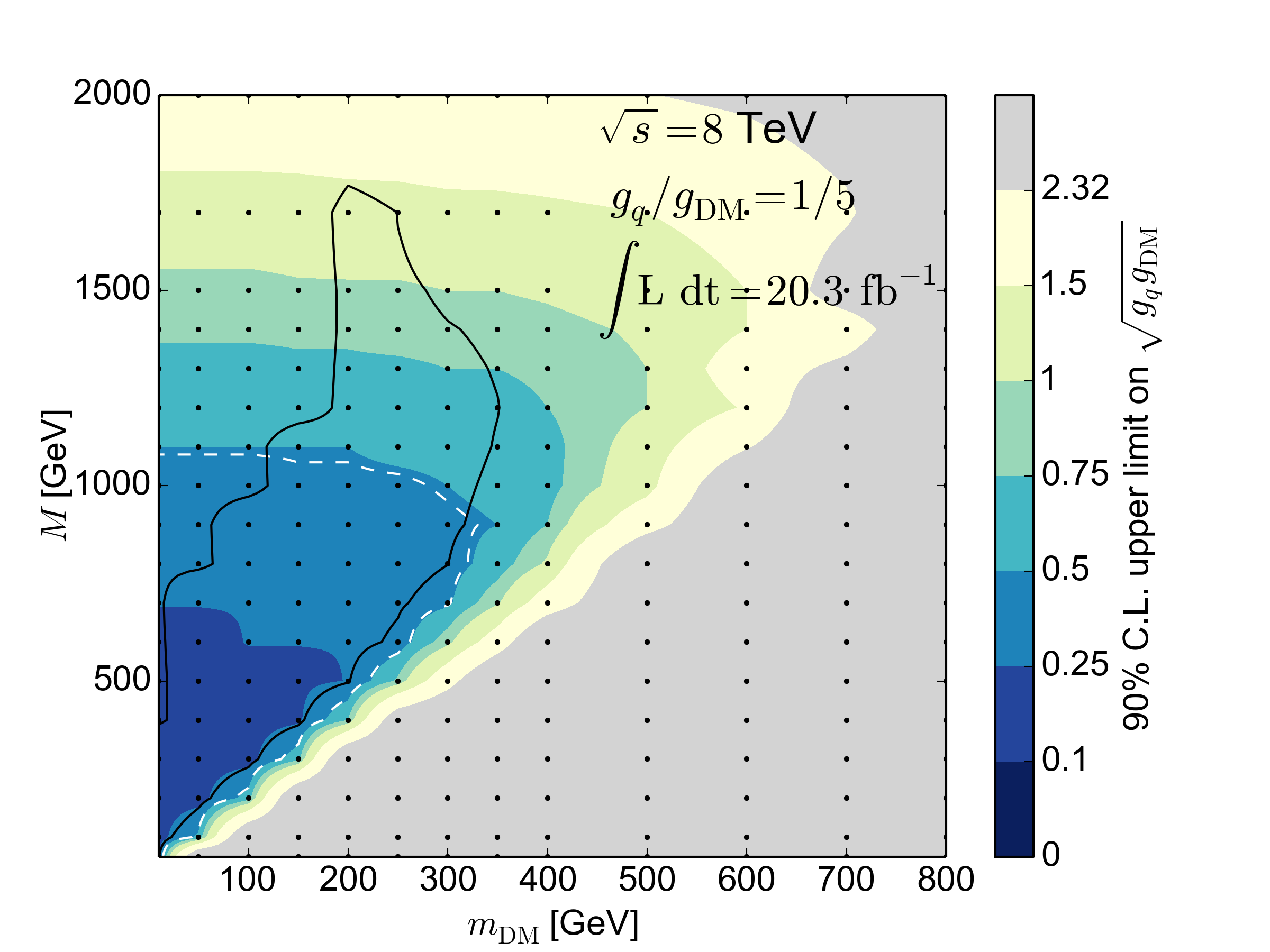}
\includegraphics[width=0.49\textwidth]{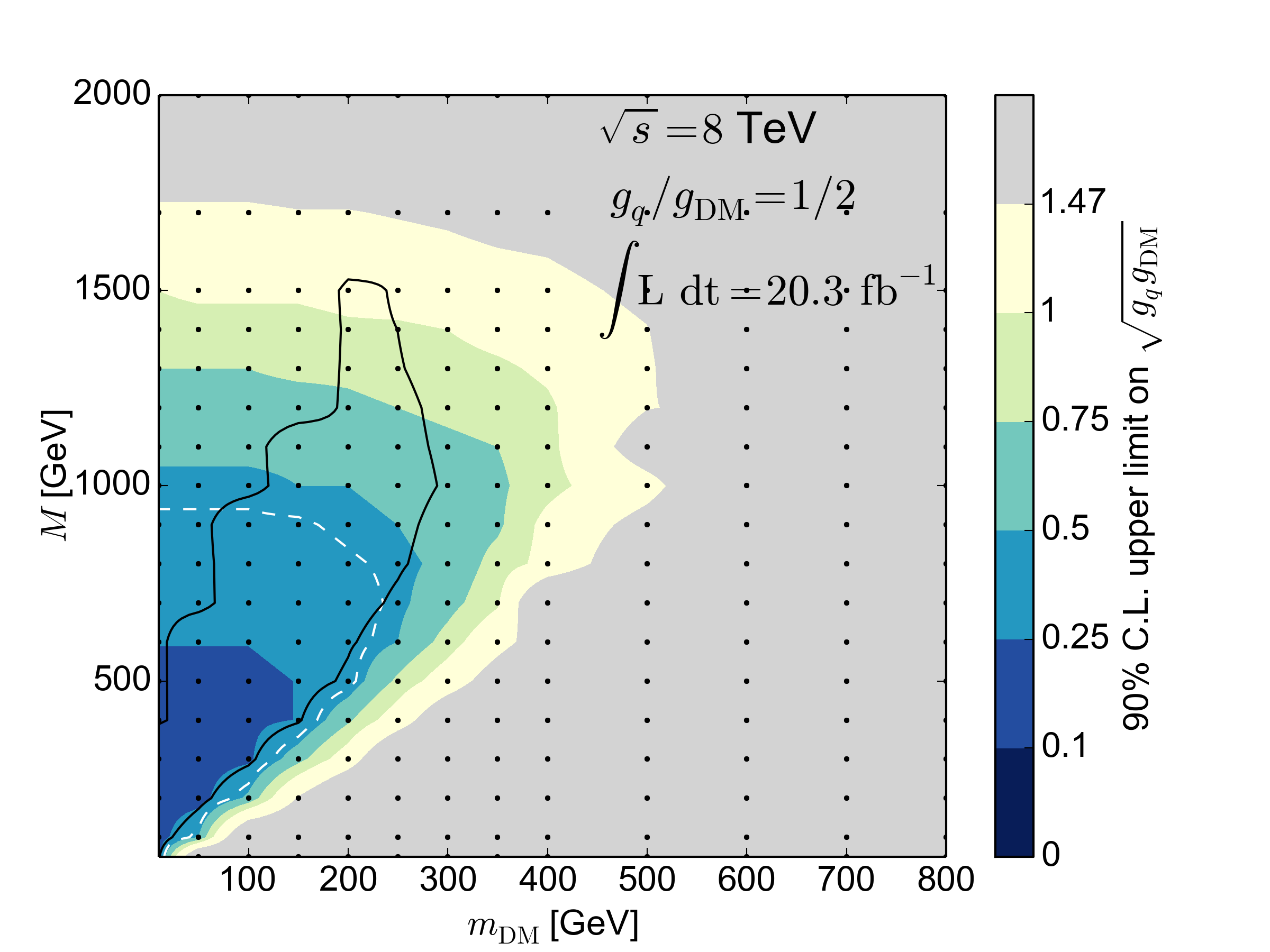}\\
\vspace{0.01\textheight}
\includegraphics[width=0.49\textwidth]{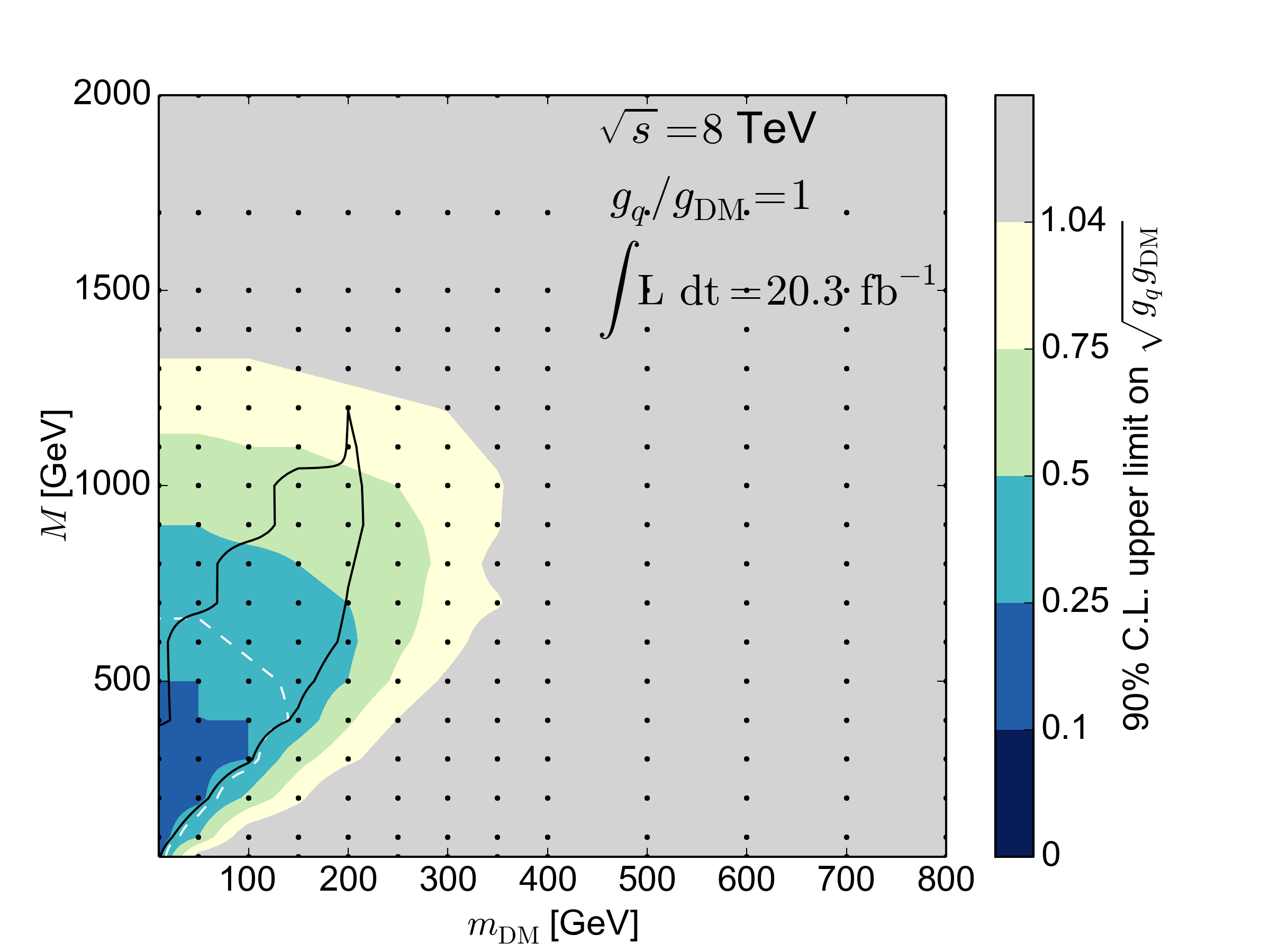}
\caption{Current ATLAS monojet constraints at 90\% C.L. on our model with the full 8 TeV dataset. The dashed white line shows where $\Gamma_\mathrm{OS}/M = 0.05$. The black dots are interpolation knots in $M - \mx$ space. The region inside the black line is naively ruled out by relic density contraints. For details see text. }
\label{full8tev}
\end{figure}

\subsection{Discussion of constraints\label{8tevdisc}}
The results of our full scan are shown in Figure~\ref{full8tev}. They cover more parameter space and are typically stronger than previous constraints using CMS results, although our limits tend to be slightly weaker in the off-shell region which can be traced back to our very conservative mediator shape reweighting. We also make an attempt at properly mapping out the limits on the $g_q/\gx < 1$ part of parameter space for the first time. We see that we can exclude mediator masses up to $\sim 1.3$ TeV for $g_q = \gx = 1$ and $\sim 0.9$ TeV for $g_q = \gx = 0.5$ at 90\% C.L.
Lower $g_q/\gx$ values are relatively well-constrained by monojet searches as expected due to resonant enhancement of the cross section. Additionally because $g_q$ is small, dijet constraints are relatively weak for this part of parameter space as we will discuss below. Although briefly studied in \cite{Buchmueller:2014yoa,An:2012va}, to our knowledge these results are the first to map out the limits in this part of parameter space in detail with the full dependence on $M$ and $\mx$.

 The area inside the black line in Figure~\ref{full8tev} indicates the region where the constraint on $\sqrt{g_q g_{\rm DM}}$ is stronger than the coupling strength which gives the correct relic abundance. Naively, $M -\mx$ values inside the region outlined by the black line would therefore lead to a larger relic abundance than observed. However, if the DM is not produced thermally or if the DM has other annihilation channels not considered here (\emph{e.g.} to leptons), this constraint is relaxed. Therefore this contour does not strictly rule out any region, but rather indicates that the `simplest' models of DM are expected to lie outside this contour. Limitations aside, this is a useful way to compare collider measurements to cosmological observations, and can be elegantly implemented in a simplified model context as we show here.

\subsection{Limits from dijet resonance searches}
Since our $Z'$ couples to quarks one could attempt to make use of limits from dijet resonance searches \cite{Aad:2014aqa,Chatrchyan:2013qha,Aaltonen:2008dn,Khachatryan:2015sja} to further constrain the model as done in for example \cite{An:2012va}. This certainly gives much stronger constraints for the $g_q/\gx > 1$ part of our parameter space which is why we don't study this part of parameter space (we show results for $g_q/\gx = 2$ in our validation study in Appendix~\ref{approximation}). For lower ratios of couplings we need to be careful since dijet resonance searches generally assume narrow resonances and in this part of parameter space the dark sector branching starts contributing to the width considerably. The dashed white line on the plots show where the width of the mediator becomes narrow enough to potentially violate such constraints assuming no additional dark sector decays (we take this to be $\Gamma_\mathrm{OS} / M \lesssim 0.05$ to be conservative, but note that there are recent searches \cite{Khachatryan:2015sja} which have constrained much wider resonances). Comparing to the detailed $Z'$ dijet analysis in \cite{Dobrescu:2013cmh} and the recent ATLAS update in \cite{Aad:2014aqa} and assuming the results won't change drastically when using an axial-vector coupling compared to a vector one, we see (note the difference of a factor of 6 in the definition of $g_q$) that the values of $g_q$ for which $\Gamma_\mathrm{OS} / M \lesssim 0.05$ for $g_q/\gx = 1/2, 1/5$ in our model typically are smaller than the values currently constrained by dijet searches, but $g_q/\gx = 1$ might be better constrained by dijet searches in the part of parameter space inside the dashed white line. Realistically the dijet resonance constraints will apply for much wider resonances than we allow here once you perform a proper analysis instead of relying on constraints set assuming narrow widths, as done in \cite{An:2012va} at parton level which suggests that monojet searches give the strongest constraints for light mediators but dijet searches are more constraining for heavy mediators even for low values of $g_q/\gx$. Due to the sensitive dependence on the width it is however worth stressing that since the model we assume here has no additional dark sector or standard model decays for the mediator, constraints set on dark matter mediators with dijet resonance searches in this part of parameter space can not be considered conservative: the width we use is the minimum width assuming equal coupling to each generation of quarks, and small changes to the dark sector can make a large difference to this width. This problem is not as pronounced when $g_q/\gx > 1$ since the width then is dominated by SM decays, which further motivates using dijet constraints over monojet ones in this part of parameter space. We also note that since the width can be large, interference effects with $Z/\gamma^*$ should be properly taken into account when using dijet searches to constrain these models -- we expect interference to play a similar role as in Drell-Yan \cite{Accomando:2013sfa} and have checked that this appears to be the case but a detailed analysis is outside the scope of this paper.

It is also possible to make use of dijet angular distributions which are sensitive to wider resonances than the dijet mass spectrum \cite{ATLAS:2012pu,Khachatryan:2014cja} and therefore can be considered more robust than dijet resonance contraints. As shown in the parton level study in \cite{An:2012va} these can also be competetive with resonance searches in some parts of parameter space, but we won't consider them further here.

\section{14 TeV Predictions}
\label{14tev}
\begin{figure}[ht]
\centering
\includegraphics[width=0.49\textwidth]{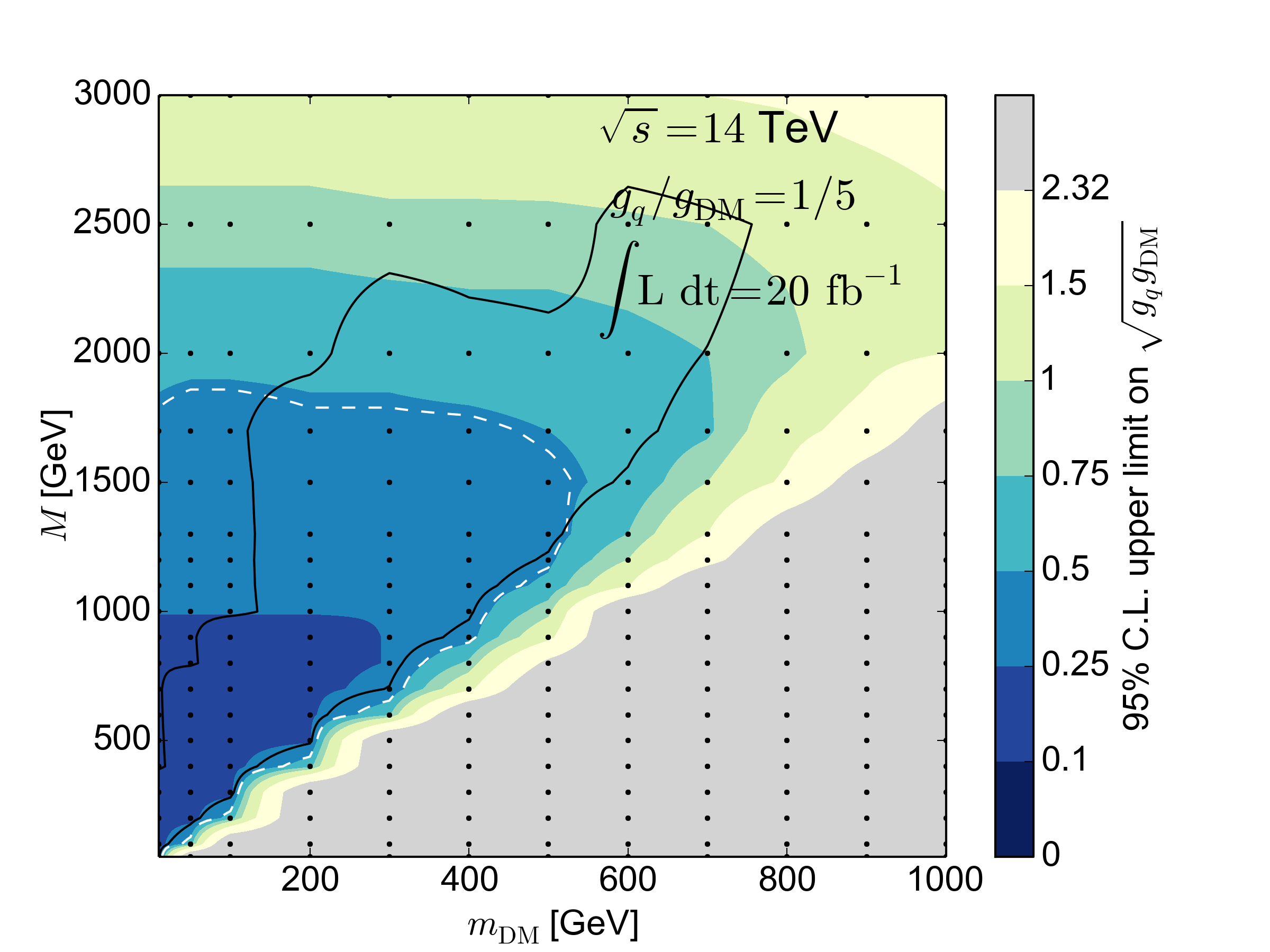}
\includegraphics[width=0.49\textwidth]{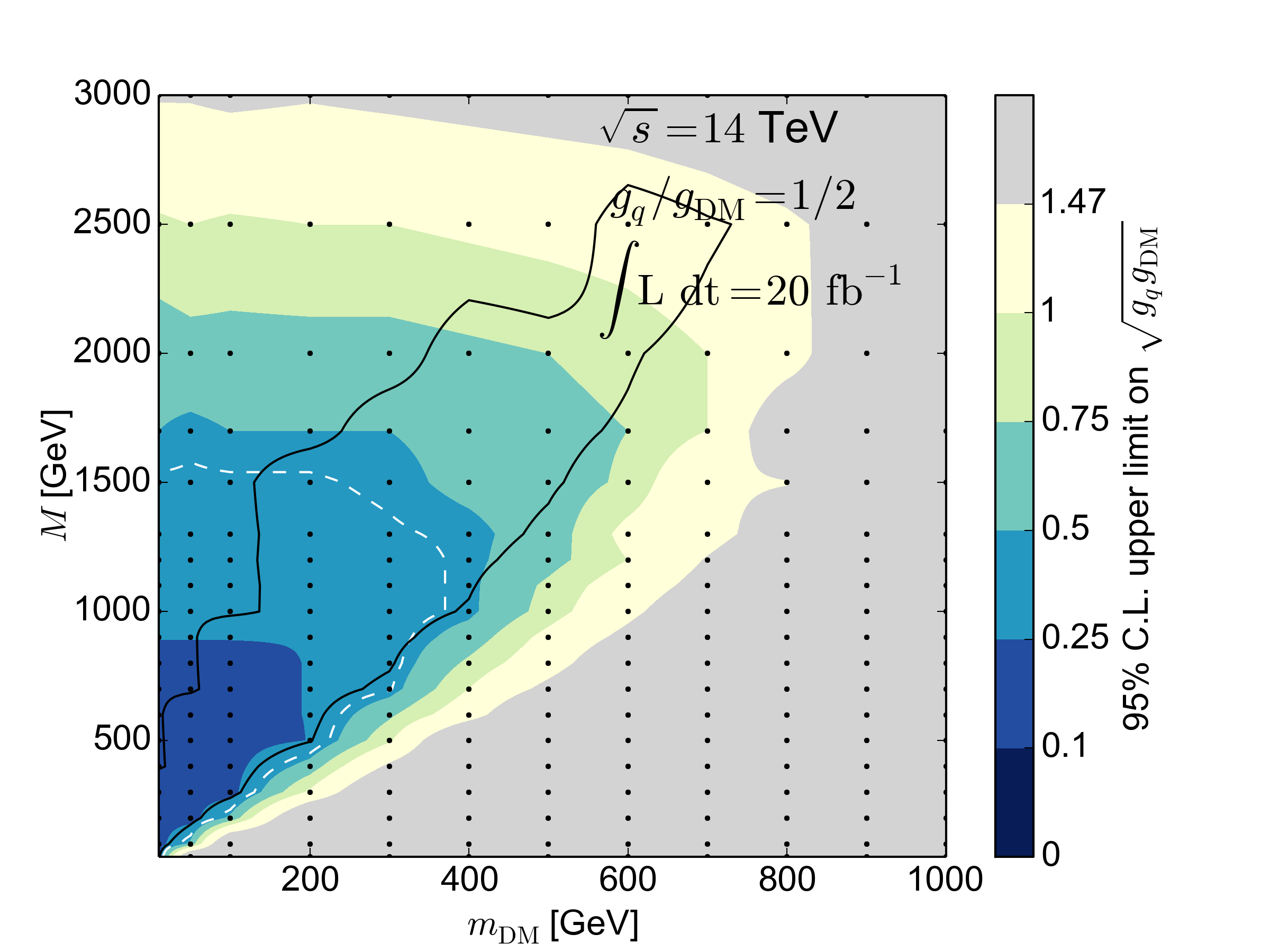}\\
\vspace{0.01\textheight}
\includegraphics[width=0.49\textwidth]{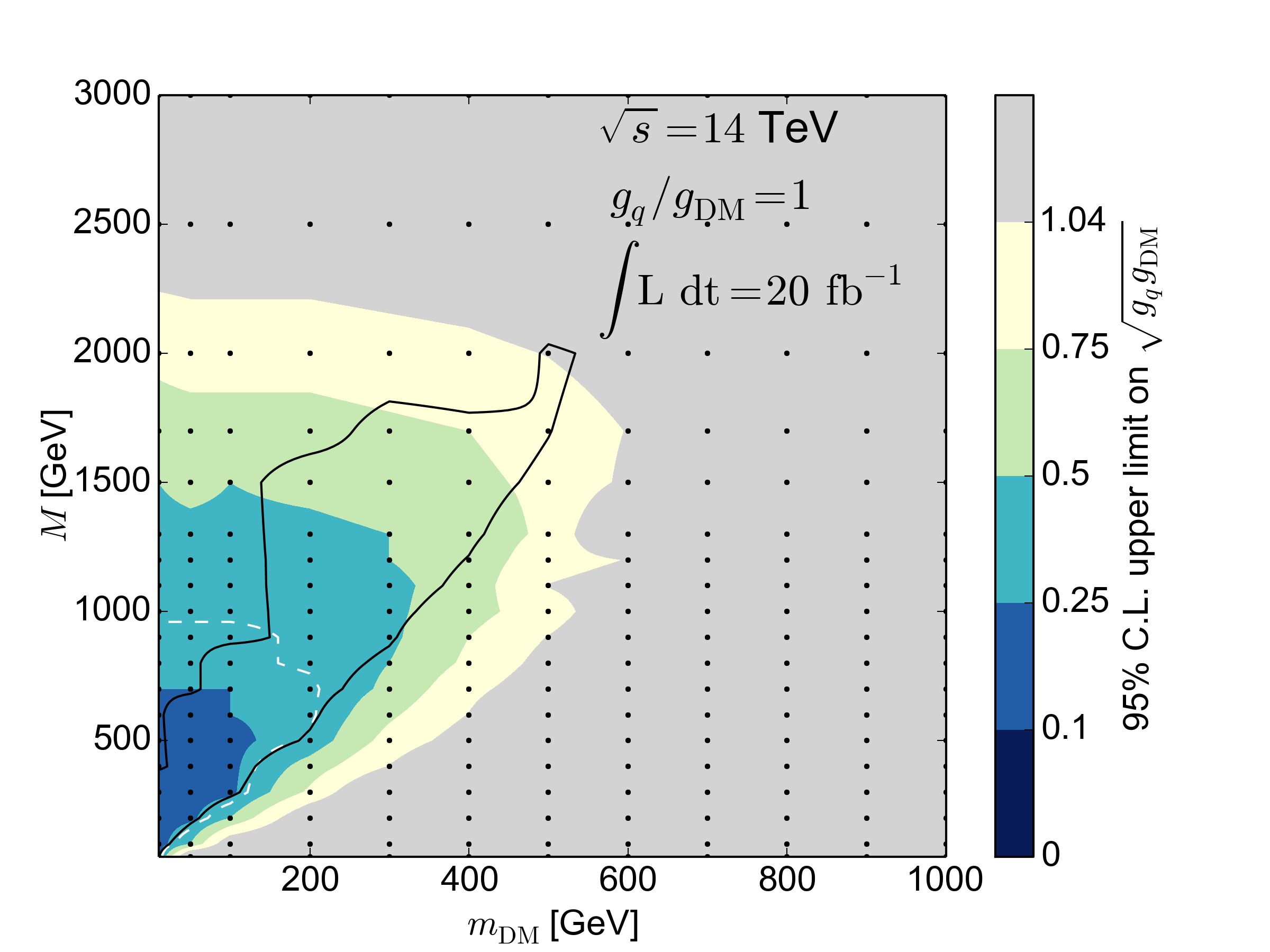}
\caption{Predictions for the reach of the ATLAS experiment at 95\% C.L. with 20fb$^{-1}$ of 14 TeV data with $\mu = 60$. The dashed white line shows where $\Gamma_\mathrm{OS}/M = 0.05$. The black dots are interpolation knots in $M - \mx$ space. The region inside the black line could naively be ruled out by relic density contraints. For details see text.}
\label{14tevpreds}
\end{figure}

We make use of the public results in \cite{ATL-PHYS-PUB-2014-007} as estimations for the expected backgrounds and hence expected cross section limits at 14 TeV with 20 fb$^{-1}$ of data from the ATLAS detector assuming a low average number of pile-up collisions ($\mu = 60$)\footnote{We thank David Salek for providing us with the exact numbers for the background estimates and expected limits.}. The details of the simulation and analysis are largely the same so we assume the detector performance will not degrade which is well-motivated under the current upgrade predictions, but our analysis is changed to mirror that in \cite{ATL-PHYS-PUB-2014-007}: we use a constant leading jet $p_T$ cut of 300 GeV and the signal regions have been redefined as detailed in Table~\ref{14tevtable}. The estimated reach with 20 fb$^{-1}$ of 14 TeV data for $g_q/\gx = 1/5, 1/2, 1$ is presented in Figure~\ref{14tevpreds}. The black line again indicates the correct relic density as discussed in Section~\ref{8tevdisc}.

\begin{table}[h]
\centering
\begin{tabular}{|c|c|c|c|}
\hline
Signal Region & SR1 & SR2 & SR3 \\
\hline
$E_{\mathrm{T}}^{\mathrm{miss}}$ $>$ [GeV]& 400 & 600  & 800 \\
ATLAS $\sigma_\textrm{vis}^\textrm{exp. 95\% CL}$ [fb] & 28 & 4.5 & 1.5 \\
\hline
\end{tabular}
\caption{Signal region definitions in the 14 TeV analysis and expected 95 \% CL exclusion limits on the visible cross section from BSM contributions. }
\label{14tevtable}
\end{table}

\section{Conclusion\label{conclusion}}

As the LHC  approaches Run II there is a clear move towards supplementing EFT analyses with simplified models, as a stronger and more robust way to constrain the dark sector. These same arguments apply to Run I data, and thus it is useful to reinterpret existing constraints on the dark sector in the simplified model framework. This has the added benefit of allowing clearer benchmarks and comparisons for future studies of simplified models at higher LHC energies and luminosities. We have demonstrated this with constraints on a simple $Z'$ model, with an axial-vector coupling. 
The parameter space for simplified models spans a minimum of 4 dimensions, making the parameter scan and visualisation of the subsequent constraints more challenging than for EFTs. The common restriction to 2-D slices of parameter space does allow for easy comparison between several constraints, but reduces our knowledge of the  model as a whole. Here we instead scan over the full 4-D parameter space, presenting results as contours, allowing us to retain the maximum information possible on constraints of a dark sector. By making some well-motivated approximations we can perform such scans in an accurate and computationally reasonable way.

This allows for a more complete understanding of the strengths of the monojet channel for constraining dark sectors and facilitates comparison to contraints from other experiments and astronomical observations, as shown here by comparing to dijet and relic density constraints. Whilst the scope of this analysis is limited to a single simplified model, this technique shows good prospects for the (re)interpretation of constraints across a broader model-space.

\section*{Acknowledgements\label{acknowledgements}}
 KN would like to thank Caterina Doglioni and Andreas Weiler for invaluable help and supervision during the summer project which this work derives from. We also thank Amelia Brennan, Sofia Vallecorsa, Stefan Prestel, Johanna Gramling, Ruth P\"ottgen, Steven Schramm, Christoph Englert, Antonio Riotto, David Miller, and Emil \"Ohman for useful discussions, help, and comments on an earlier version of this manuscript, and Emanuele Re for pointing out a mistake in the discussion of previous results. 
 
 KN thanks the University of Glasgow for a College of Science \& Engineering scholarship, CERN for a summer studentship, and the DM@LHC '14 organizers for financial support which made this work possible.

\appendix
\section{Validation of Procedure using EFT Limits \label{validation}}

We use the ATLAS monojet analysis for 10.5 fb$^{-1}$ of 8 TeV data \cite{ATLAS-CONF-2012-147} to validate our procedure and approximations, so we apply the following cuts: We require at most two jets with $p_\mathrm{T} > 30$ GeV and $|\eta| < 4.5$, with $|\eta^{j1}| < 2$ and $\Delta \phi (j2, E_\mathrm{T}^\mathrm{miss}) > 0.5$ where $j1$ and $j2$ are the leading and subleading jet respectively. We define four signal regions based on $p_\mathrm{T}^{j1}$ and $E_\mathrm{T}^\mathrm{miss}$, given in Table \ref{regions}.
\begin{table}[h]
\centering
\begin{tabular}{|c|c|c|c|c|}
\hline
Signal Region & SR1 & SR2 & SR3 & SR4 \\
\hline
$p_\mathrm{T}^{j1}$ \& $E_{\mathrm{T}}^{\mathrm{miss}}$ $>$ [GeV]& 120 & 220  & 350  & 500 \\
ATLAS $\sigma_\textrm{vis}^\textrm{95\% CL}$ [fb] & 2800 & 160 & 50 & 20 \\
\hline
\end{tabular}
\caption{Signal region definitions in the 10.5 fb$^{-1}$ 8 TeV analysis and ATLAS 95 \% CL exclusion limits on the visible cross section from BSM contributions.}
\label{regions}
\end{table}

Our overall limit-setting procedure has been validated by recreating the ATLAS limits set on $\Lambda$ for the D8 EFT operator which corresponds to our simplified model. A comparison for SR3 is presented in Table~\ref{lambdaLimits}. We consistently overestimate the limit by a few percent which reflects the less advanced nature of our detector simulation, however the agreement is good enough for our purposes as we have sub-2\% differences for $\mx$ values which are relevant for us. Note that we only perform the comparison for SR3 as it usually is the most discerning signal region and the only one for which ATLAS results are reported, however we assume the results are similar for the other signal regions. Similarly we assume this agreement carries over to our analyses of the full 8 TeV dataset and our 14 TeV predictions, which is well motivated since the full dataset 8 TeV analysis was conducted under similar conditions and due to the stated ATLAS upgrade goals for the upcoming higher energy LHC run respectively.

\begin{table}[h]
\centering
\begin{tabular}{|c|c|c|c|}
\hline
$\mx$ [GeV] & ATLAS 95\% CL on $\Lambda$ [GeV] & Our 95\% CL on $\Lambda$ [GeV] & Difference [\%] \\
\hline
 $\le$80 & 687 & 700 & +1.9  \\
 400 & 515 &  525 & +1.9 \\
 1000 & 240 & 250 & +4.2  \\
\hline
\end{tabular}
\caption{Comparison of limits set on the D8 EFT operator by ATLAS \cite{ATLAS-CONF-2012-147} and us using only SR3.}
\label{lambdaLimits}
\end{table}

\section{Validation of Cross Section Reweighting \label{reweighting}}

\begin{figure}[ht]
\centering
\includegraphics[width=0.49\textwidth]{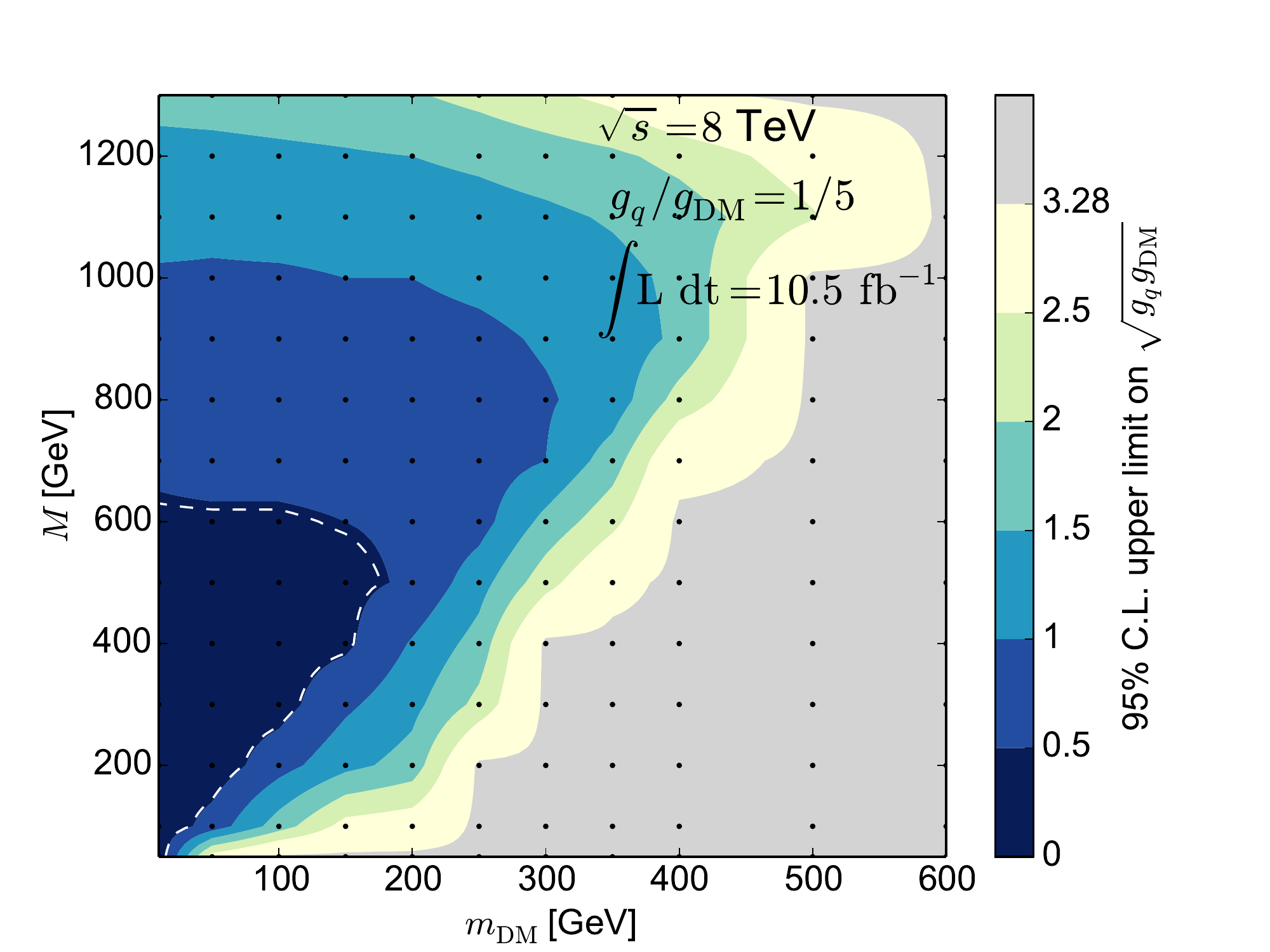}
\includegraphics[width=0.49\textwidth]{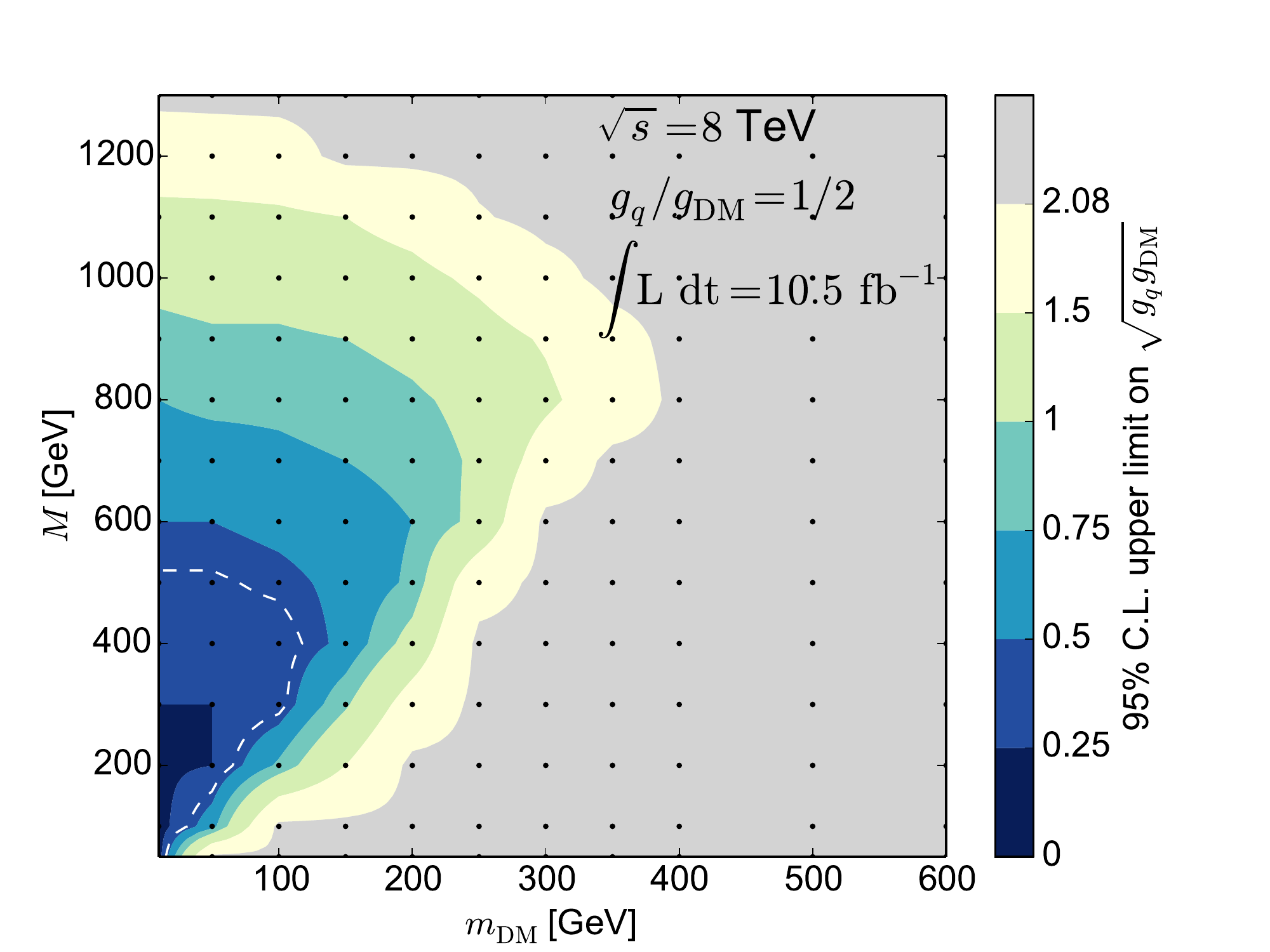}\\
\vspace{0.02\textheight}
\includegraphics[width=0.49\textwidth]{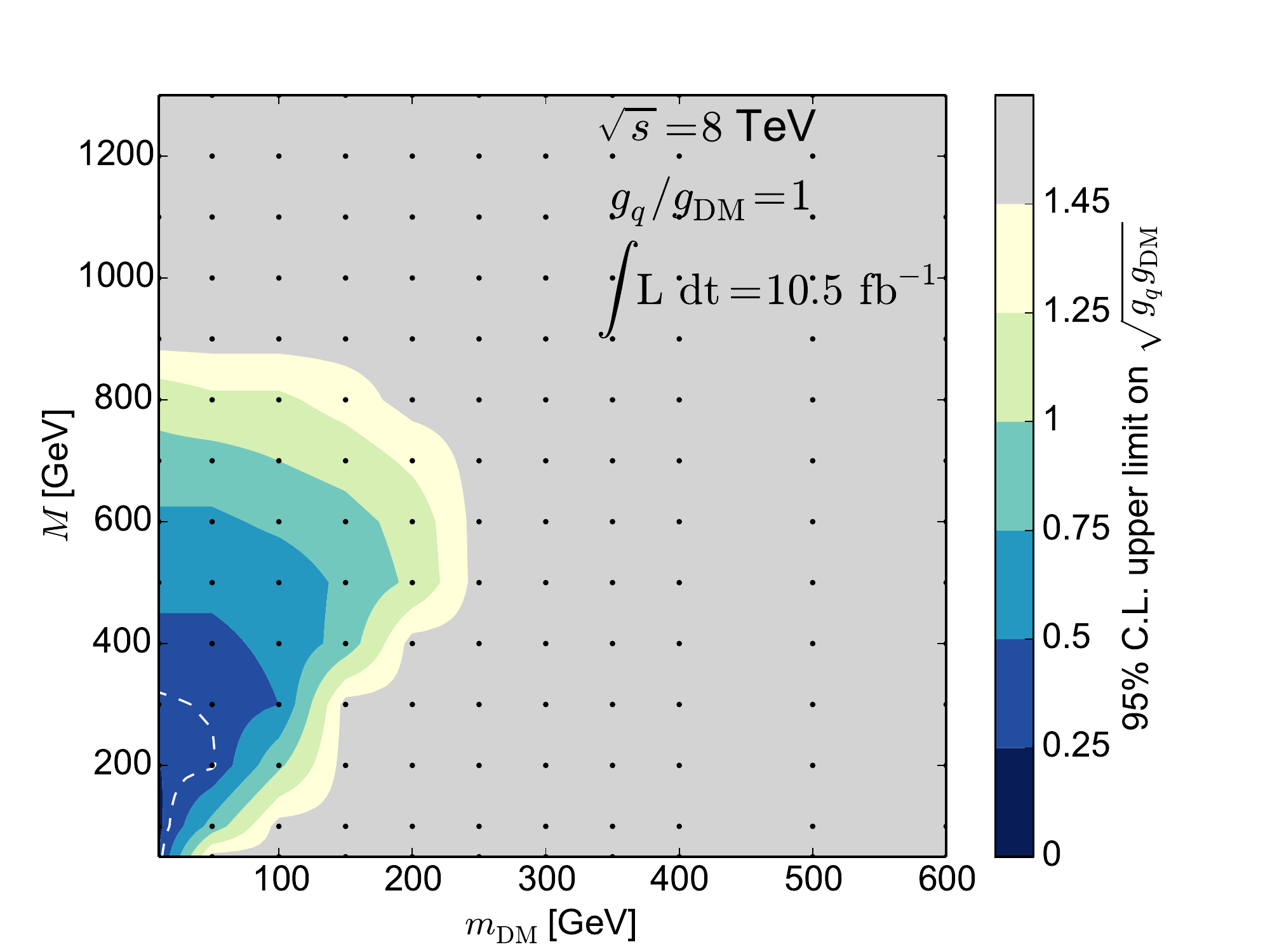}
\includegraphics[width=0.49\textwidth]{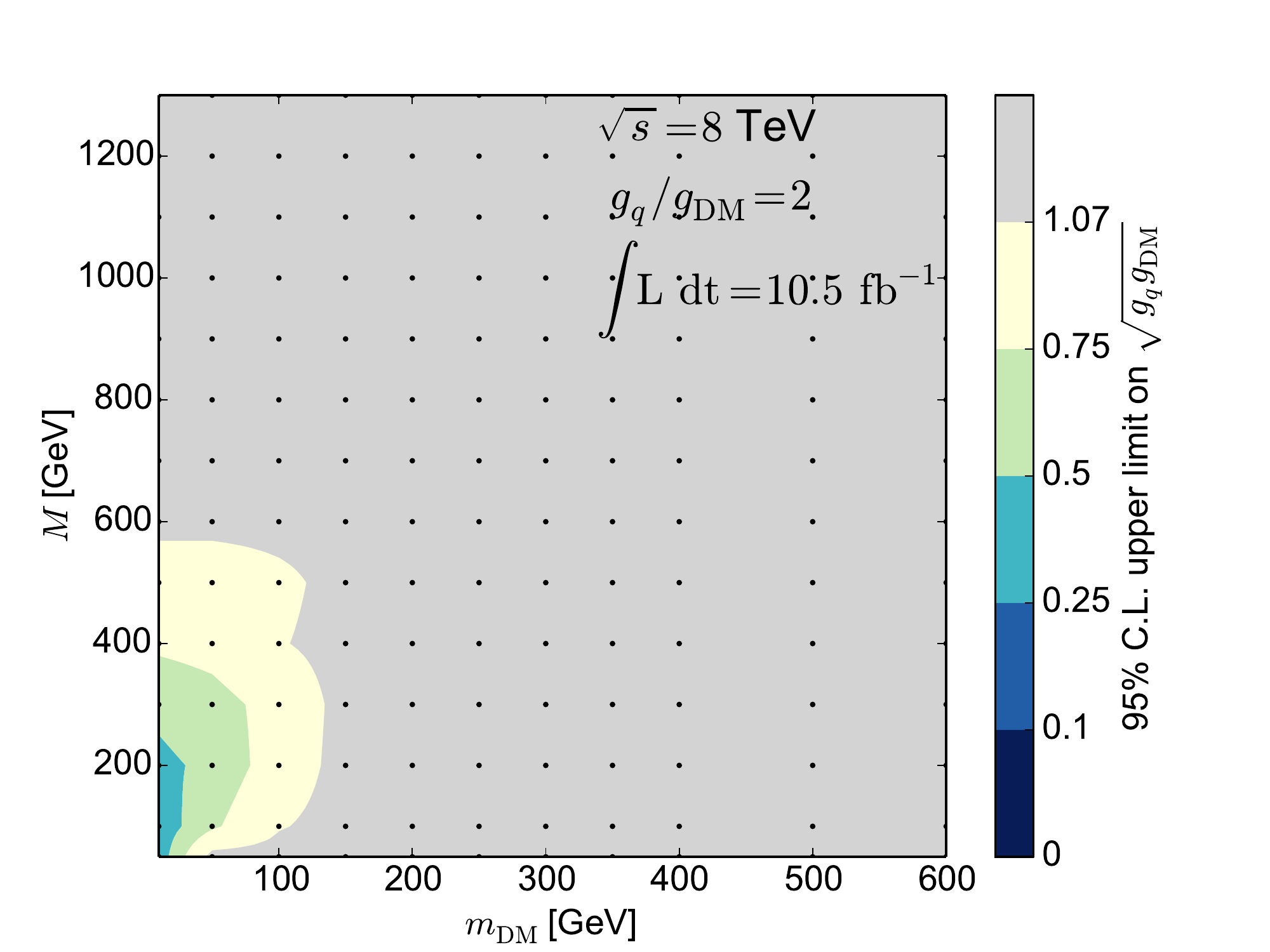}\\
\caption{Our results using interpolation in $M - \mx - \gx \cdot g_q$ space. The dashed white line shows where $\Gamma_\mathrm{OS}/M = 0.05$. The black dots are interpolation knots in $M - \mx$ space. See the text for further details. }
\label{fullresults}
\end{figure}

\begin{figure}[ht]
\centering
\includegraphics[width=0.49\textwidth]{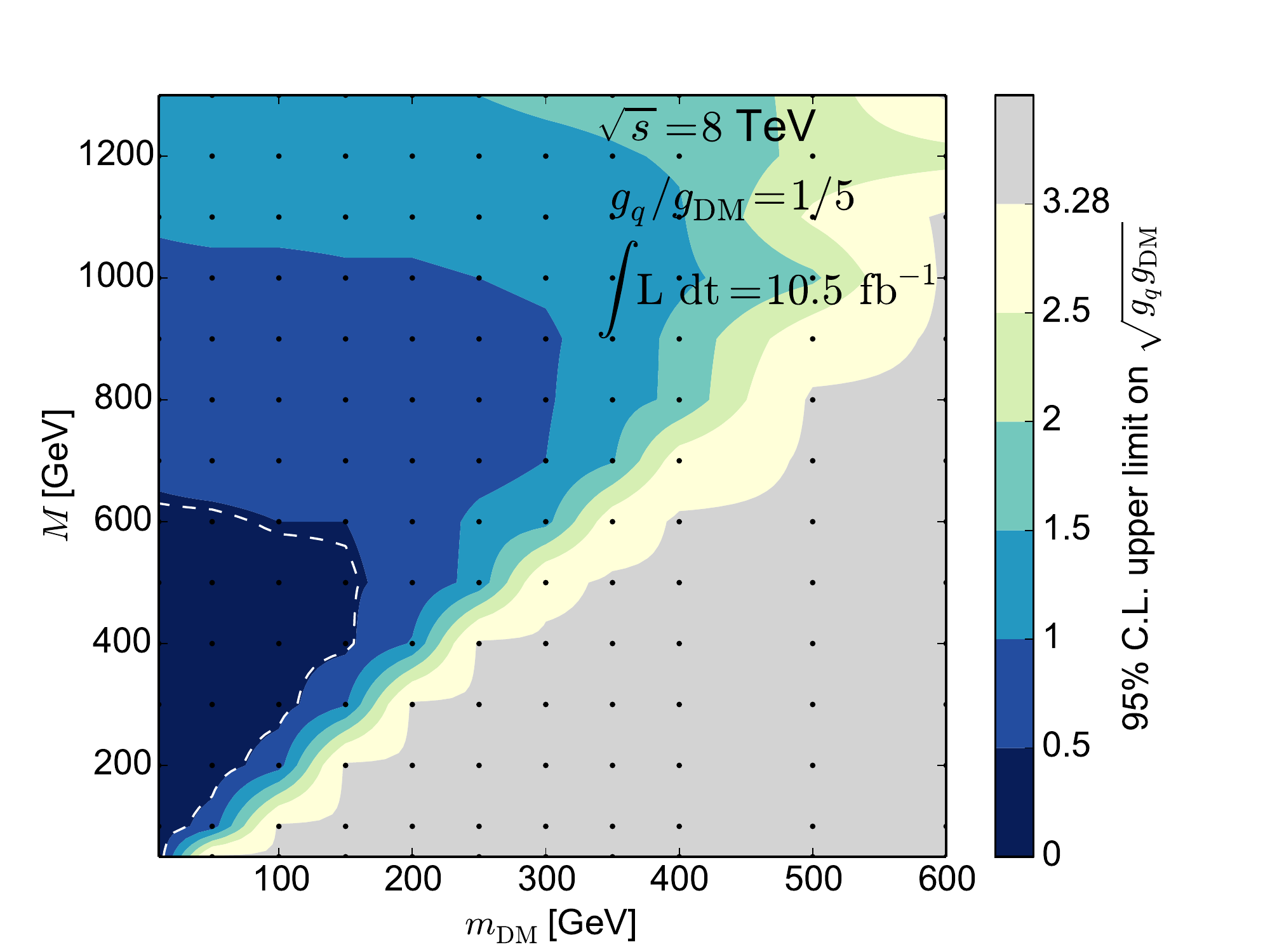}
\includegraphics[width=0.49\textwidth]{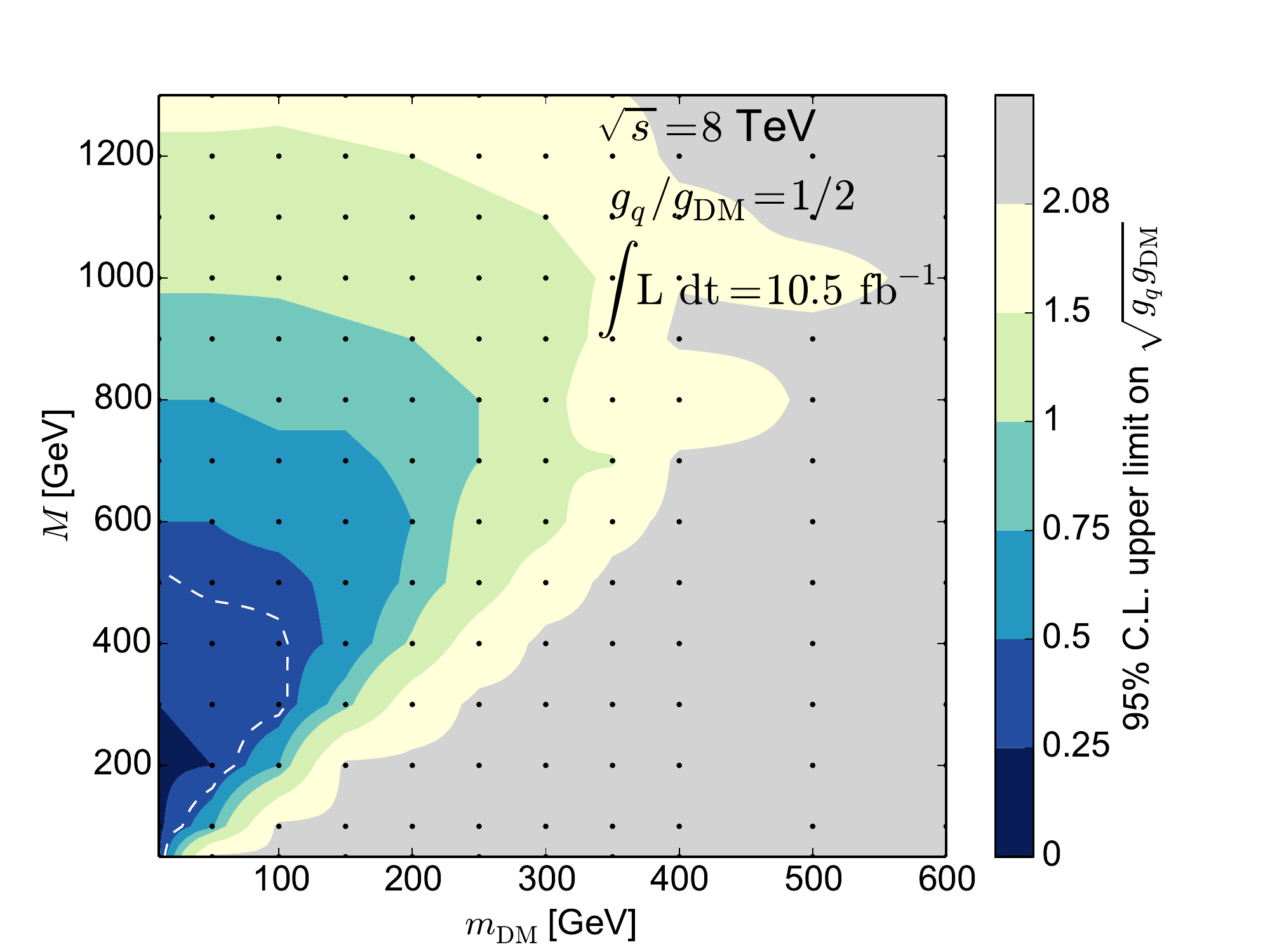}\\
\vspace{0.02\textheight}
\includegraphics[width=0.49\textwidth]{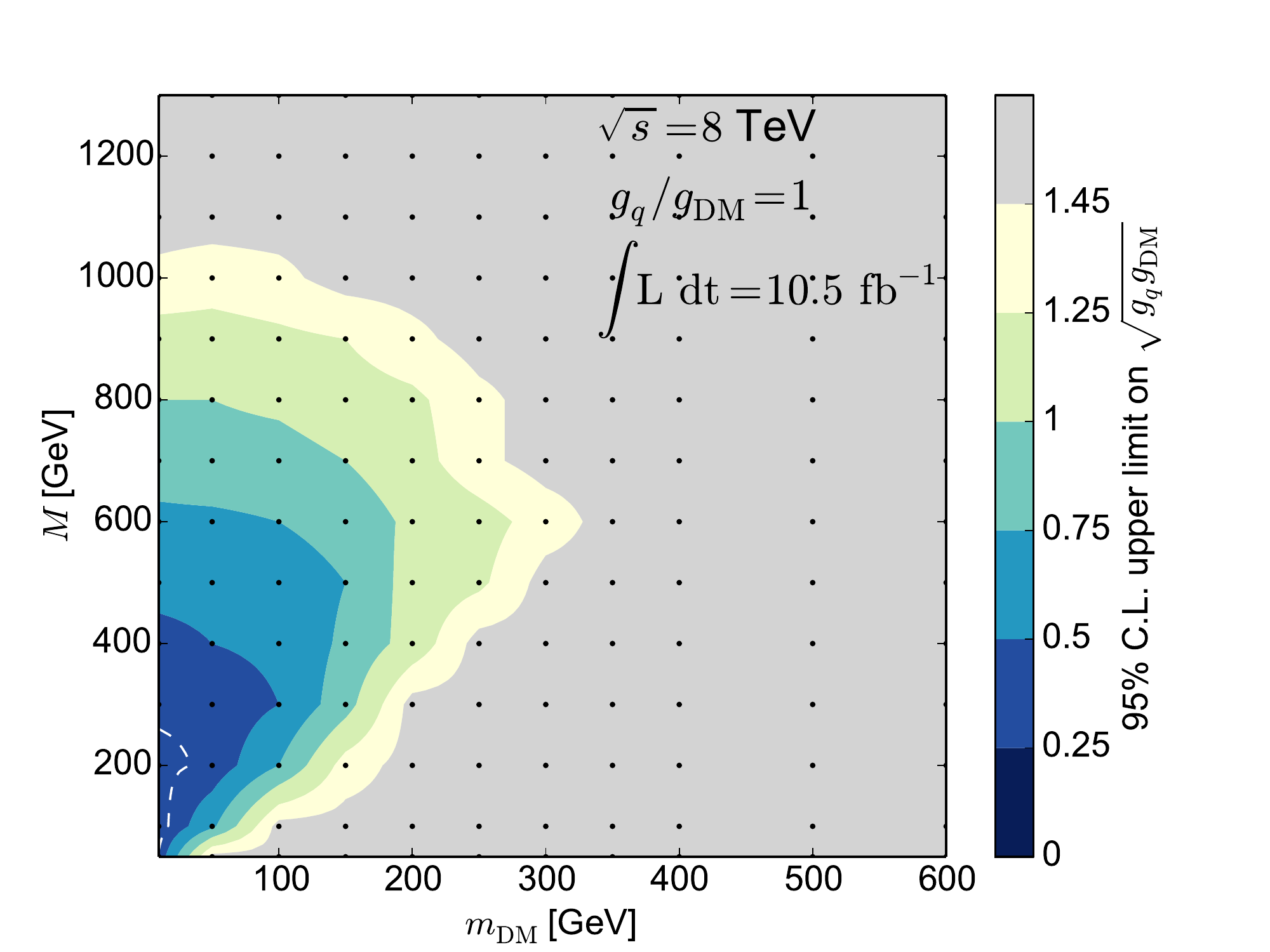}
\includegraphics[width=0.49\textwidth]{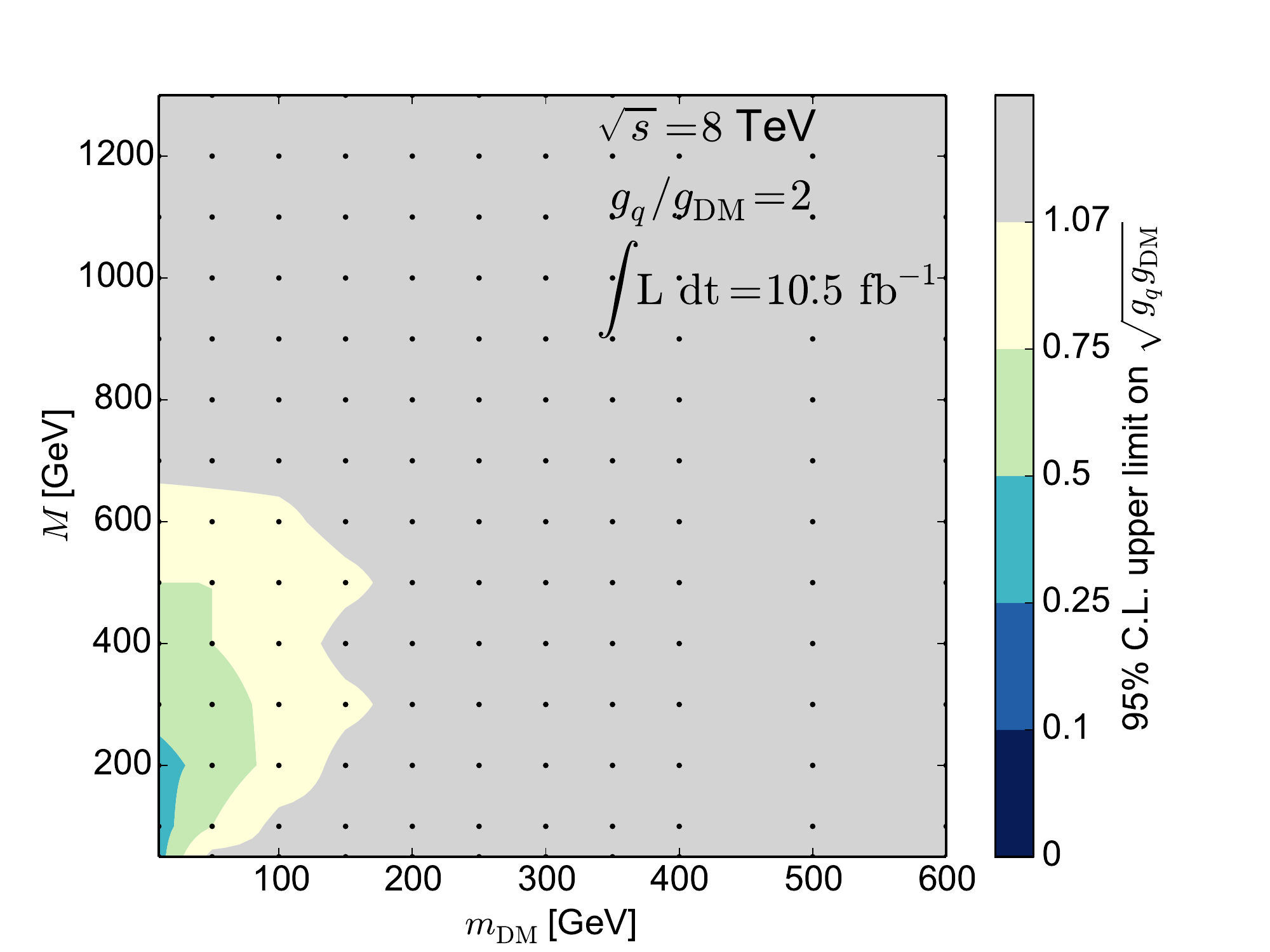}\\
\caption{Our results using interpolation in $M - \mx$ space and the cross section approximations detailed in the text. The dashed white line shows where $\Gamma_\mathrm{OS}/M = 0.05$. The black dots are interpolation knots in $M - \mx$ space. See the text for further details.}
\label{reweighted}
\end{figure}

\begin{figure}[ht]
\centering
\includegraphics[width=0.49\textwidth]{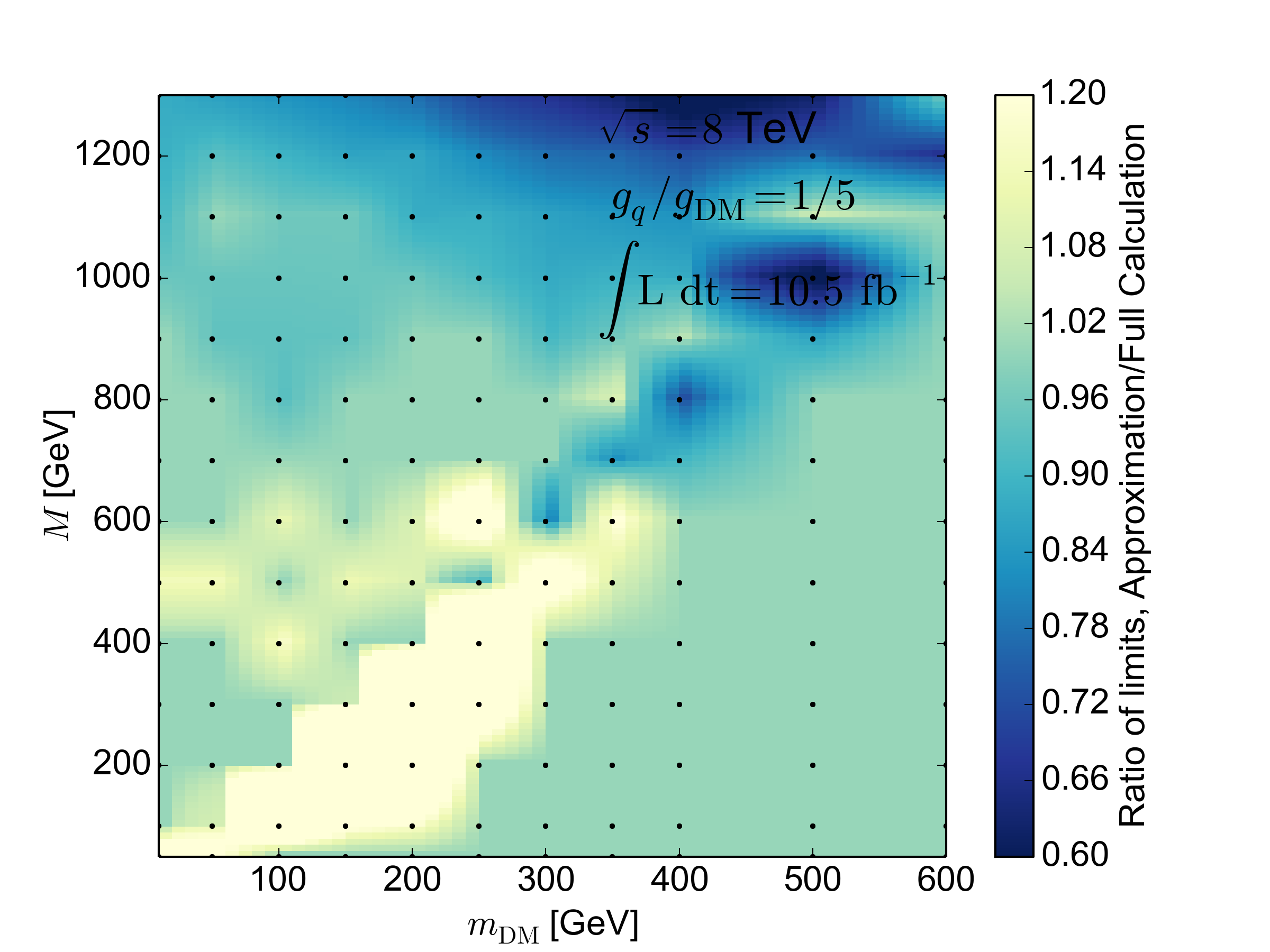}
\includegraphics[width=0.49\textwidth]{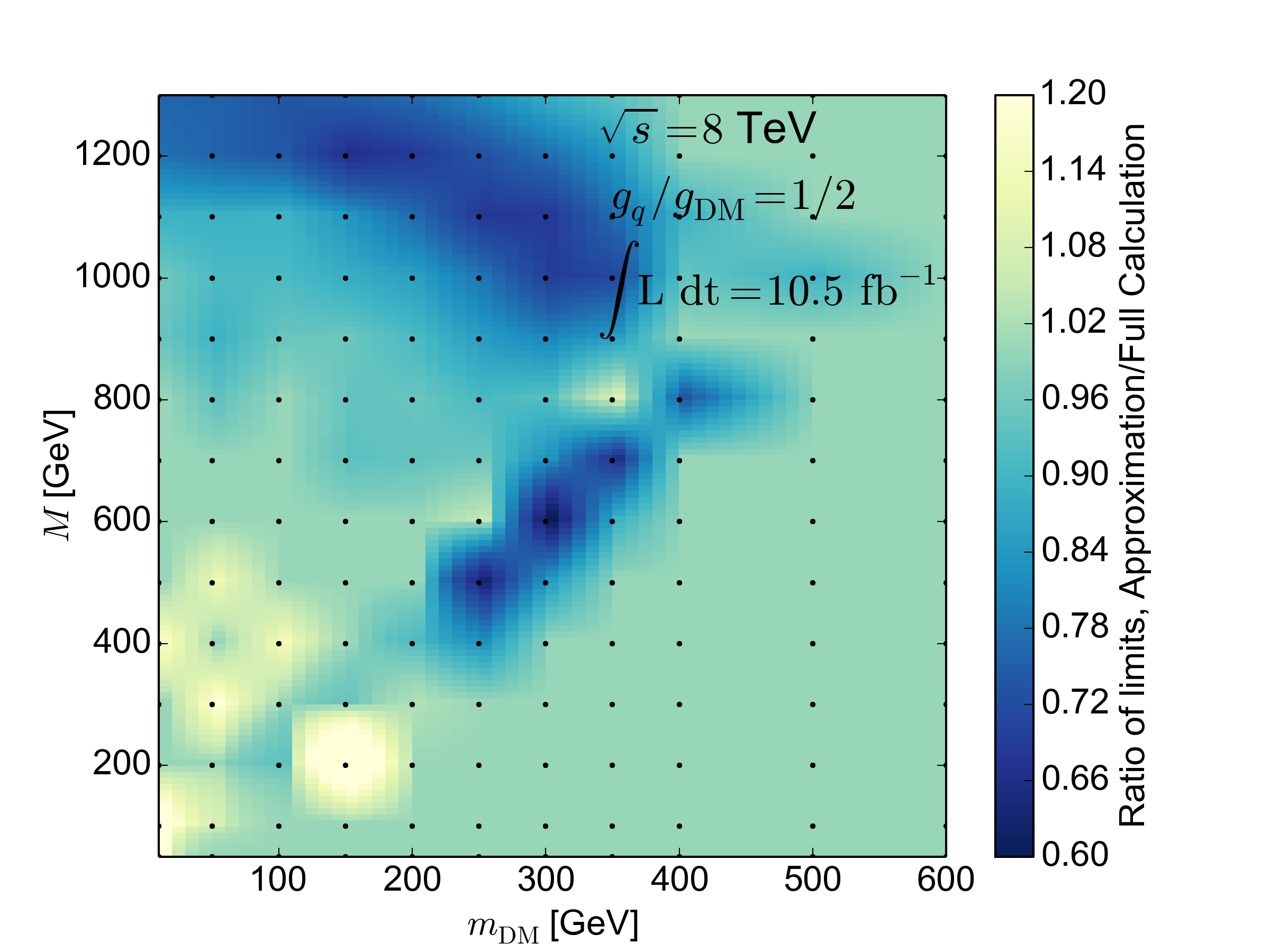}\\
\vspace{0.01\textheight}
\includegraphics[width=0.49\textwidth]{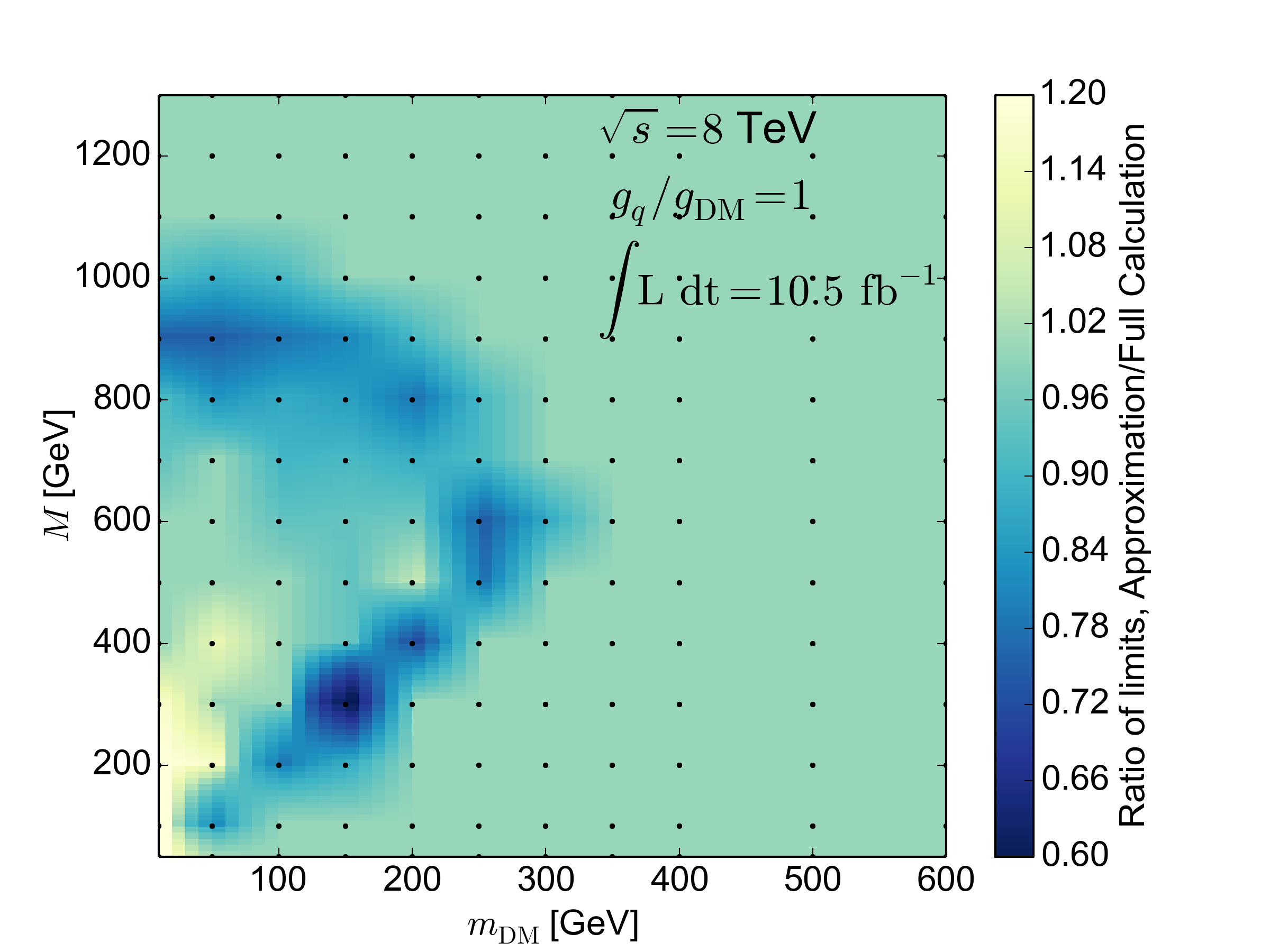}
\includegraphics[width=0.49\textwidth]{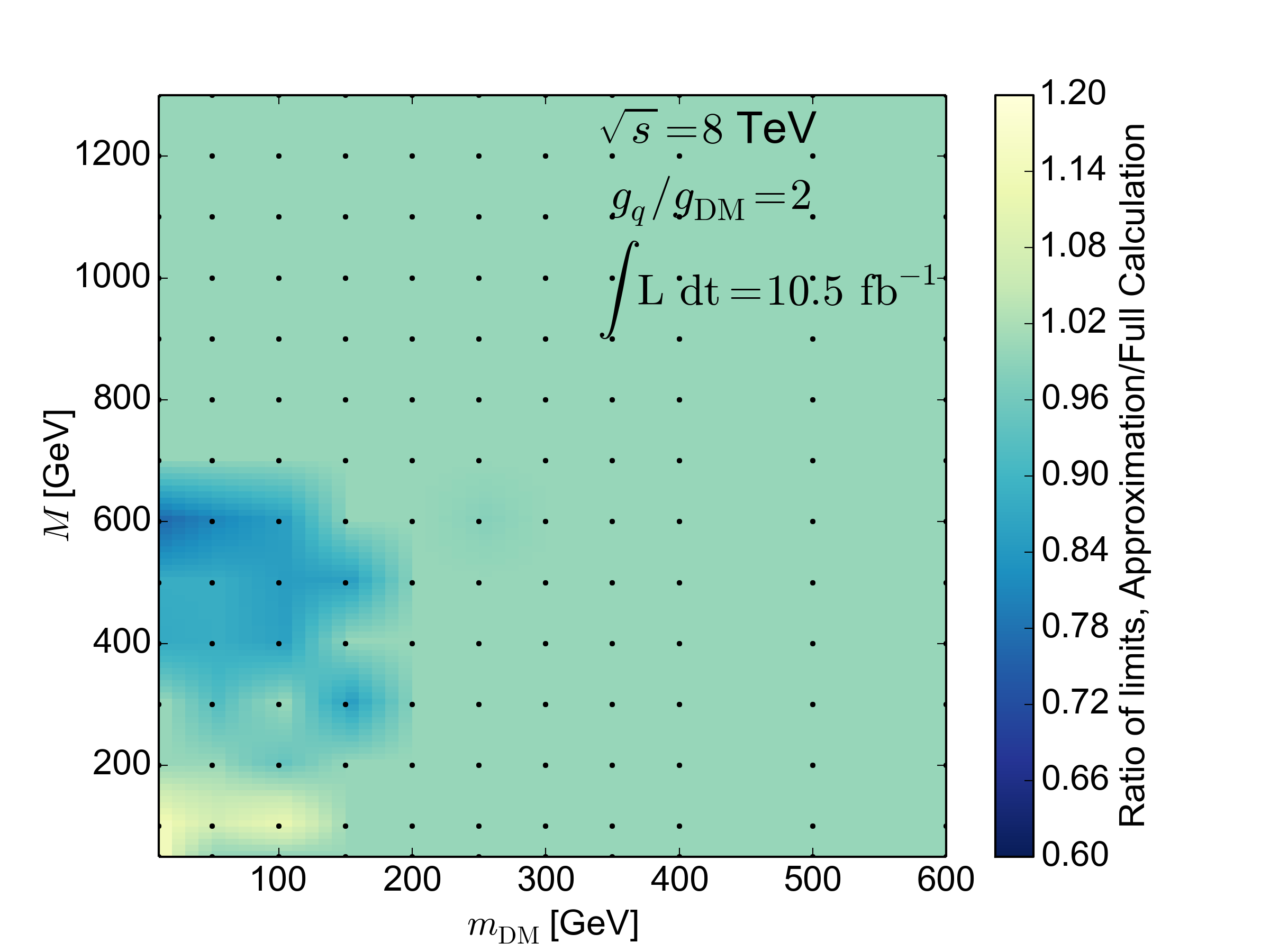}\\
\caption{Ratio of the results using interpolation in $M - \mx$ space and the cross section approximation detailed in the text to using a full interpolation in  $M - \mx - \gx \cdot g_q$ space. The cross section approximations are conservative in the bright yellow (light) areas, and overestimates the limit in the dark blue (dark) areas. The black dots are interpolation knots in $M - \mx$ space. Note the ratio takes values higher than 1.2 in some parts of parameter space but the colourbar is restricted since the approximations are conservative there. See the text for further details. }
\label{ratio}
\end{figure}

Our limits set using results from Ref.~\cite{ATLAS-CONF-2012-147} using interpolation in $M - \mx - \gx \cdot g_q$ are presented in Figure~\ref{fullresults}, limits set using the cross section approximation including the width mentioned in Section~\ref{approximation} are presented in Figure~\ref{reweighted}, and the ratios of the limits set in the two cases are presented in Figure~\ref{ratio}. To visualise the breakdown of our approximations we extend the limit of our parameter space to $\Gamma_\mathrm{OS}/M < 1$.

Values of $g_q/\gx > 1$ are hardly probed at all by monojet searches as evident from our results for $g_q/\gx = 2$: such models are much better constrained by dijet searches which motives not including these in our main study.
%

 

\subsection{Using a cross section approximation including the width\label{approximation}}

We compare our results to ones obtained by reweighting the cross section for a single value of $\gx \cdot g_q$ to see how well the simple cross section approximations: 
\begin{equation}
 \sigma \propto
\left\{
	\begin{array}{ll}
		g_q^2 \gx^2/\Gamma_\mathrm{OS}  & \mbox{if } M > 2\mx \\
		 g_q^2 \gx^2 & \mbox{if } M < 2\mx
	\end{array}
\right.
\label{reweight}
\end{equation}
reproduce the full results. Additionally we perform a separate reweighting to correct for the Breit-Wigner shape of the propagator as for the full results, although only before finding the limit using \ref{reweight}\footnote{This could be improved by using an iterative approach but we avoid doing so here to keep the procedure simple.}. The results using this reweighting are presented in figure \ref{reweighted} and the ratio of the limits in Figure~\ref{ratio}. As expected they work well for sufficiently small values of $\gx \cdot g_q$ that $\Gamma_\mathrm{OS} \ll M$. As $\Gamma_\mathrm{OS}/M \to 1$ the cross section approximations become inaccurate and the less accurate shape reweighting becomes an issue. This motivates us to use these cross section approximations for our study of the full 8 TeV ATLAS dataset and our 14 TeV predictions while remaining robust by further limiting the parameter space to $\Gamma_\mathrm{OS}/M < 0.5$.

\section{Comparison with Relic Density Constraints\label{relics}}

If dark matter was produced thermally in the early universe, there is a simple relationship between the thermally averaged dark matter self-annihilation cross section $\langle\sigma v\rangle_{\rm ann}$, and the observed relic abundance $\Omega_{\rm DM}h^2$. For a given model, this allows us to find the coupling strength which provides the correct relic abundance as a function of $(m_\chi, M)$. 
This scenario is by no means a certainty; If the observed dark matter was produced through some mechanism other than thermal production, or if some new physics has an effect on the connection between the self-annihilation rate and the abundance at freezeout, this relationship breaks down. At the same time, thermal dark matter is a well-motivated scenario, and is a useful way to get a sense of the regions of parameter space within which we expect the gravitationally-observed DM to lie. 

The observed relic abundance can be approximated as \cite{Jungman:1995df}
\begin{equation}
\Omega_{\rm DM}h^2\simeq \frac{2\times2.4\times 10^{-10}\,{\rm GeV}^{-2}}{\langle\sigma v\rangle_{\rm ann}}.
\label{simplerelic}
\end{equation}
Combined with Planck constraints of $\Omega_{\rm DM}^{\rm obs}h^2=0.1199\pm0.0027$ \cite{Ade:2013zuv}, we see that $\langle\sigma v\rangle_{\rm ann}\simeq 4.0\times 10^{-9}\,{\rm GeV}^{-2}$ for thermal relic DM. 
Rather than using this approximation, we use the well-known  formalism described in, for example, Refs.~\cite{Gondolo:1990dk,Bertone:2004pz}
to constrain $\langle\sigma v\rangle_{\rm ann}$ by simultaneously solving an expression for the freezeout temperature as a function of $\langle\sigma v\rangle_{\rm ann}$, and the relic abundance as a function of both $\langle\sigma v\rangle_{\rm ann}$ and the freezeout temperature.  This technique ceases to be valid near resonance and when the Breit-Wigner expression for the width breaks down. We therefore restrict our calculations to regions where $\Gamma/M < 1$ and $M > \mx$. We use the annihilation rate to quarks for our model as calculated in Ref.~\cite{Busoni:2014gta}.

We indicate on Figures~\ref{full8tev} and \ref{14tevpreds} a contour, within which the LHC constraint on the coupling is stronger than the coupling  which would give thermal relic DM. As mentioned in the text, for regions inside this line, the coupling strength is constrained to be less than the coupling which gives the correct relic density. Hence, the annihilation rate is smaller than required, and the relic density will naively be too large. For DM to lie in this region, either the thermal relic scenario must break down, or the DM annihilates via additional channels not considered here.

\bibliographystyle{JHEP}
\bibliography{bibfile}

\end{document}